\documentclass[aps,prd,notitlepage,groupaddress,onecolumn,%
tightenlines,11pt,eqsecnum,longbibliography]{revtex4-2}

\usepackage[a4paper,margin=25mm]{geometry}
\usepackage{amsmath,amssymb,mathtools,bm}
\usepackage{siunitx}
\usepackage{booktabs,array,tabularx,multirow}
\usepackage{graphicx}

\usepackage{hyperref}
\usepackage{xcolor}
\usepackage{enumitem}

\usepackage{maybemath}

\hypersetup{colorlinks=true,%
linkcolor=blue!45!black,%
citecolor=blue!45!black,%
urlcolor=blue!45!black,%
pdftitle={Space--Based Lock--In Measurement of the Gravitational Constant}}

\newcommand{\Cov}{\operatorname{Cov}}
\newcommand{\Var}{\operatorname{Var}}

\newcommand{\dd}{\mathrm{d}}
\newcommand{\ee}{\mathrm{e}}
\newcommand{\ii}{\mathrm{i}}

\newcommand{\DC}{\mathrm{DC}}

\newcommand{\mathbm}{\boldsymbol}

\begin{document}

\title{Space--Based Lock--In Determination of 
Newton's Gravitational Constant}

% Engineered Geometry, Covariance Analysis and Mission Parameters}

%\title{Engineered Gravitational Lock--In Metrology in Space: \\
%A Dual-Channel Experiment for Measuring Newton's Gravitational Constant}

\author{Ulrich D. Jentschura}
\affiliation{Department of Physics and LAMOR, 
Missouri University of Science and
Technology, Rolla, Missouri 65409, USA}

\begin{abstract}
Inspired by recent ground-breaking space-based measurements
of Newton's gravitational constant, we outline a proposal for 
a general layout of an optimized space-based approach 
that combines the advantages of a largely disturbance-free
environment with lock-in techniques, null coordinates and 
the method of covariances. The method of covariances, 
and cross-covariances, is used for an optimized 
statistical analysis of data originating from two distinct 
experimental channels. One of two distinct measurement channels relies on an
approximately constant acceleration in the gravitational field of static sources
masses, while the second channel involves a lock-in measurement based on test
and source masses with oscillating coordinates. 
Conceivable mission parameters are analyzed
in terms of the optimum scale for the source masses and launch capabilities 
of currently available space launch systems. A candidate location
for a largely disturbance-free position in geostationary orbit is identified.
\end{abstract}

\maketitle

\tableofcontents

%
% Introduction
%
\section{Introduction}
\label{sec1}

%
% Metrological Background
%
\subsection{Metrological Background}
\label{sec1A}

Measurements of Newton's gravitational constant $G$
have a long history~\cite{Gillies1987Index,Gillies1997Review}.
According to the most recent adjustment~\cite{MohrEtAl2025} 
of the fundamental constants
recommended by the Committee on Data of the International Science Council (CODATA),
$G$ occupies quite a unique position among the 
constants of nature. It carries
the largest relative uncertainty ($2.2 \times 10^{-5}$)
of all physical constants in the 
CODATA adjustment. Furthermore, one encounters 
further complications and inconsistencies in
conjunction with precision
measurements of $G$; namely, the experimental data for $G$ 
require a large expansion factor 
of 3.9 applied to the uncertainties of the input data
in order to reduce the maximum normalized residual of the 
16~input data to a value of $2.0$
(see Table XXX of Ref.~\cite{MohrEtAl2025}).
Here, we recall that the 
normalized residual is the difference between a individual result
for $G$ and the adjusted value of $G$,
divided by the estimated standard deviation.
This reduction of the maximum normalized residual 
of the input data by the application of expansion factors 
is routinely applied by the CODATA Task Group on
Fundamental Constants~\cite{MohrEtAl2025,Tiesinga2021CODATA},
in the case of input data with internal discrepancies. 
For comparison, data on the high-precision 
frequencies in simple atomic systems
used in the CODATA adjustment
receive a much smaller expansion factor of 1.7
(see Tables XI, XII, and XIV of Ref.~\cite{MohrEtAl2025}),
while data for the fine-structure constant
receive an expansion factor of 2.5
(see Table XXV of Ref.~\cite{MohrEtAl2025}).
From a metrological point of view,
the low accuracy of the current CODATA 
value of $G$ and the large expansion factor 
applied to the input data for $G$ 
lead to a very dissatisfactory situation,
especially when compared to the progress
recently achieved in the 
redefinition of the International
System of Units (Syst\`{e}me International, SI)
in terms of the Quantum SI, with the 
latter being described in the 
Special CODATA adjustment of 2017 (see Ref.~\cite{Mohr2017CODATA}),
in which the Boltzmann constant $k_B$ 
and the Planck constant $h$ were redefined
to assume numerical values without uncertainty.
The gravitational constant $G$ is very far
from this status, and a resolution of the 
discrepancies would solve an important 
physics problem. The persistent disagreement
among nominally precise measurements has repeatedly motivated calls
for conceptually distinct experiments, improved environmental
stability, and continued investigation rather than abandonment of the
problem~\cite{Quinn2000MeasuringBigG,Quinn2014DontStopBigG}.

Compounding the situation, 
we observe that the geocentric gravitational
coefficient $\mu_\oplus = G \, M_\oplus$ 
of the Earth is 
$\mu_\oplus = 398600.4419(2) \, {\rm km}^3/{\rm S}^2$,
with an uncertainty of $5 \times 10^{-10}$
(see Ref.~\cite{Dunn1999}).
This value is not in agreement with the 
value of $\mu_\oplus = 398600.435507 \, {\rm km}^3/{\rm S}^2$,
which is given without an uncertainty 
estimate in Table~2 of Ref.~\cite{Park2021}.
However, the difference between the two 
recent determinations of $\mu_\oplus$ is 
only $1.6 \times 10^{-8}$ and thus 
smaller than the uncertainty of the current
CODATA value for $G$ by a factor of roughly $1.3\times 10^3$.
The mass of the Earth,
measured in terms of its ``gravitational coupling
strength'' $\mu_\oplus$, is much better known than
the gravitational constant $G$ itself.

\subsection{Earth--Basis versus Space--Basis}

Newton's law for two point masses,
\begin{equation}
F=G \, \frac{m_1 \, m_2}{r^2},
\end{equation}
looks elementary, but a laboratory determination of $G$ requires an absolute
realization of force, mass, length, and geometry in the presence of
environmental perturbations. 
The gravitational constant
must be inferred from the weak attraction of
macroscopic source masses whose density distributions, positions, and
mechanical couplings must be known with extraordinary accuracy. 
It is almost ironic that torsion balance methods,
which were already used in the 1790s 
(see Ref.~\cite{Cavendish1798}),
are still being used (in an improved form)
up to this day and lead to competitive accuracy 
in some of the most recent 
experiments~\cite{LiEtAl2018}.
For terrestrial experiments,
the central metrological difficulty is not raw signal size alone
(one might argue that the signal size could always
be increased on the basis of larger source masses~\cite{REMARK}). 

Recently, a ground-breaking experiment~\cite{Armano2019GSpace}
reported on the first space-based determination 
of $G$, using two independent methods.
The first displaced one of the two nominal LISA Pathfinder 
test masses and inferred the change in gravitational attraction from 
electrostatic suspension forces. The second used the slow loss 
of nitrogen from the cold-gas system as a time-dependent 
source-mass signal in the ordinary drag-free 
differential-acceleration observable. Using the two methods,
the obtained values in the two channels read
$G_1 =(6.71\pm0.42)\times10^{-11}\,\mathrm{m^3\,kg^{-1}\,s^{-2}}$ and
$G_2 =(6.15\pm0.35)\times10^{-11}\,\mathrm{m^3\,kg^{-1}\,s^{-2}}$.
The corresponding relative uncertainties are $6.3\%$ and $5.6\%$.
The measurement described in Ref.~\cite{Armano2019GSpace}
constitutes a ground-breaking, 
yet, in terms of accuracy, proof-of-principle principle measurement
because it was based on the reuse of an apparatus whose mass distribution,
electrostatic-control mode, propellant metrology, and calibration had
been optimized for the LISA Pathfinder 
mission rather than for a measurement of $G$. 

The present paper asks about possible optimizations of
the space-based approach for a drastic improvement of 
the precision. It is known that a number of 
systematic effects
in high-precision experiments, such as for critical exponents (MISTE), are
reduced in microgravity environments~\cite{BaHaLiDu2007}. 
However, one should be careful in regard
to exaggerated hopes for an immediate drastic improvement of experimental
accuracy.  A space-based experiment does not abolish systematic effects, but
rather, replaces terrestrial disturbances with a different set:
Among these, we find spacecraft gravity, orbital tides, and
control-system forces. The attraction is that a test mass can remain
unsupported for long intervals, while displacement is measured
interferometrically. The experiment can therefore exploit long free-evolution
times and coherent narrow-band detection. The success of LISA Pathfinder
Mission, which was designed to demonstrate the 
feasibility of low-frequency gravitational wave detection,
established that free-fall quality and differential acceleration sensing at the
femto-$\mathrm{m\,s^{-2}/\sqrt{Hz}}$ level are physically realizable in flight
(see Refs.~\cite{Armano2016PRL,Armano2018PRL,Armano2024Actuation}). 

We here advocate a proposal that combines the advantages of the space-based
environment with lock-in methods. Recently, such lock-in methods have been
proposed in other contexts where the separation of the experimental signal from
background effects presents considerable challenges, namely, a for the
detection of laser-induced vacuum birefringence (see
Ref.~\cite{BuJeYo2025prr}). Our proposal does not rely on the assumption that
space is disturbance-free. It relies on converting the disturbance environment
into measured auxiliary channels and on making the gravitational signal
spectrally and geometrically distinctive.

\subsection{Measurement Philosophy}
\label{philosophy}

Our experimental proposal is based on unsupported test masses during data
taking intervals which is made possible in a space-based apparatus.  For
enhanced verification, the proposal involves two independent experimental
channels.  The first of these involves static source masses and the
time-dependent determination of test mass trajectories using laser
interferometric techniques, and constitutes and analog of the Satellite Energy
Exchange (SEE) proposal in a well-controlled 
environment~\cite{SandersDeeds1992,AlexeevEtAl1994,Sanders1996CommentsSEE,%
SandersEtAl1999,SandersEtAl2000,AlexeevEtAl2000SEE,Alexeev2001SEE,%
Sanders2010SEE,Melnikov2016SEE}. 
We shall henceforth refer to the 
first channel as the direct channel (DC),
where, of course, the acronym DC bears resemblance to the 
unidirectional motion of charge (here replaced
by the unidirectional motion of a test mass) in an electromagnetic setting.
The second experimental channel involves periodic, symmetric
oscillations of the positions of the source masses, which eliminates a number
of systematic effects and enables the use of lock-in detection techniques. 
We shall henceforth refer to the 
oscillatory trajectory of source masses combined with
lock-in techniques as the oscillatory lock-in channel (LC).
In both experimental channels, a third test mass supplies a symmetry and null
channel. For an optimized determination of $G$ based on data taking in both
above-mentioned channels, the method of covariances 
(Appendix~F of Ref.~\cite{MoTa2000}) 
is available. Combining the data from
two channels with the method of covariances,
one arrives a global adjustment in
which correlations are retained. One notes that the
apparatus is overdetermined in the sense that the same $G$ must explain the
mean trajectory, the synchronous response, higher harmonics, geometry changes,
and null combinations, as well as other observables;
the overdetermination also constitutes an essential 
element used in the CODATA adjustment of fundamental 
constants in general~\cite{MoTa2000,Mohr2017CODATA,Tiesinga2021CODATA,MohrEtAl2025},

For completeness, we should include some remarks
on an alternative proposal for a 
space-based determination of Newton's gravitational constant
as described in Ref.~\cite{FeldmanEtAl2016DeepSpaceG},
where  a ``gravity clock'' was proposed in
which a retroreflector executes natural oscillations through an axial
tunnel in a layered spherical source mass.
In the latter case, $G$  is
obtained primarily from the gravitational oscillation period and the
calibrated mass distribution of the source body.
Although dynamically elegant, the gravity-clock concept places an
exacting burden on source calibration in the sense that 
the test mass moves in the
near field of a bored, composite source body whose detailed
three-dimensional mass distribution---not merely its total mass and
outer radius---determines the restoring force and hence the inferred
value of $G$. The proposal discussed here uses spherical source
and test masses.

\subsection{Significant Ingredients}

{\em Drag-free control.---}Our proposal uses, in an essential way, six ingredients
and techniques which have been developed over the last decades,
to arrive at a strategy (including metrological strategy)
for an improved determination of big--$G$.

The drag-free concept uses an internal free body as an inertial reference. The
spacecraft is commanded to follow that body using micropropulsion, thereby
shielding the body from solar radiation pressure and other non-gravitational
forces acting on the spacecraft. Early formulations date to the development of
precision relativistic satellite experiments \cite{Cannon1962DragFree}.
Gravity Probe B, GOCE, MICROSCOPE, LISA Pathfinder, and related missions
developed complementary elements of this technology.

For the present mission, drag-free control has two possible roles. In an
integrated architecture, the spacecraft follows one test mass or a suitable
common-mode combination. In the preferred detached-laboratory architecture, 
the laboratory would 
supply local shielding and metrology while a more
distant service spacecraft performs communication, navigation, and periodic
station keeping. 

{\em Satellite Energy Exchange.---}The 
Satellite Energy Exchange (SEE) proposal considered a heavy ``Shepherd''
and a lighter ``Particle'' in a long drag-free capsule. Their relative orbital
motion contains a characteristic energy-exchange or horseshoe-like encounter
from which $G$, inverse-square-law deviations, and equivalence-principle
effects can be inferred. Detailed studies examined trajectory sensitivity,
source-body quadrupole moments, orbital parameters, and charging
\cite{Gillies1997,Sanders2010SEE,Alexeev2001SEE}.

The present proposal preserves the central SEE insight: $G$ is extracted from
the dynamics of freely moving bodies rather than a static suspension. It
differs in geometry and metrological strategy. Two source masses are placed
symmetrically around three test masses, the sources are periodically
displaced, and the observables are decomposed into a mean differential channel,
a lock-in channel, and a central null channel. Thus the concept should be
described as a modulated, internally redundant extension of the SEE philosophy,
not as a rejection of it.

In the SEE proposal, one uses a massive shepherd and one or two light particles;
here, two large engineered source masses and multiple test masses
are used. While, in the SEE, 
the relative orbital encounter generates the signal,
here, it is the engineered source geometry.
The Complete nonlinear trajectory is the observable in the 
SEE, while here, differential DC and lock-in observables are measured.
In the SEE proposal, 
the trajectory is determined largely by orbital mechanics,
while here, the trajectory is prescribed by experimental design.
The extraction of $G$ is envisaged
by fitting complete encounter trajectories  in the SEE,
while here, the extraction of $G$ 
is assumed to proceed by a 
generalized least-squares adjustment of multiple complementary observables.
Internal consistency checks could be achieved 
from from repeated orbital encounters  in the SEE,
while here, one has internal consistency checks from 
independent DC, AC, null, rotated and background channels.

{\em LISA Pathfinder.---}LISA Pathfinder 
contained two nominally free Au--Pt cubic test masses housed in
gravitational reference sensors. The electrode housings provided capacitive
sensing and electrostatic actuation, while an optical interferometer measured
the relative displacement along the sensitive axis. The mission demonstrated
differential acceleration performance better than its original requirement and
close to that required for LISA over a substantial frequency range
\cite{Armano2016PRL,Armano2018PRL,Armano2017Capacitive}.

LISA will use three spacecraft linked by laser interferometry. Each spacecraft
contains free-falling test masses whose non-interferometric degrees of freedom
are controlled electrostatically; along the principal measurement direction the
objective is to avoid applied force. ESA adopted LISA in January 2024, with
launch planning in the mid-2030s \cite{ESALISA2024,LISADefinition2023}.

{\em Electrodes and Actuation.---}The 
LISA Pathfinder inertial sensor used a hollow electrode housing surrounding
a $46\,\mathrm{mm}$ test-mass cube with millimeter-scale gaps. 
Currently available mechanical actuators 
reach sub-micrometer precision \cite{Actuators}.
They are based on ball-screw and roller-screw stages.
An original approach is used in the LISA Pathfinder 
mission~\cite{Armano2024Actuation,Bassan2018Crosstalk,Weber2003GRS}.
Electrodes opposing the test masses form differential capacitors. 
A displacement $x$ of the test mass
changes the two capacitances with opposite signs.
The same electrodes can apply force.
For a voltage $V$ on an electrode pair, 
the electrostatic energy is $U=\tfrac12 C(x) \,V^2$, 
and so, for a distance-dependent capacitance, 
one can generate a mechanical force on the 
test mass equal to $\partial U/\partial x$.
LISA Pathfinder demonstrated
nano--Newton actuation while characterizing gain noise, low-frequency voltage
fluctuations, and 
cross-talk~\cite{Armano2024Actuation,Bassan2018Crosstalk,Weber2003GRS}.

{\em Charge Management.---}
Cosmic rays and energetic particles charge isolated test masses. A net charge
$q$ couples to stray electric fields and magnetic fields through Coulomb and
Lorentz forces. LISA Pathfinder used ultraviolet photoemission to transfer
charge without physical contact and measured both charging rates and
charge-induced acceleration noise in flight
\cite{Armano2017ChargeNoise,Armano2018ChargeControl,Armano2023Charging,Sumner2009Charging}.
The present experiment should inherit a contactless UV charge-management system
and schedule discharge operations between data taking or at frequencies
chosen not to overlap the gravitational modulation
so that the lock-in signal is not disturbed.

{\em Laser Interferometric Readout.---}A 
laser interferometer measures optical phase and therefore relative
displacement. It does not require a moving reference mirror in the ordinary
engineering sense; a static displacement determines a static phase, while
acceleration appears through the curvature of the phase trajectory. In the
present apparatus, however, the most useful signal is deliberately time
dependent. If, as in the proposed LC channel,
$x(t)=X\cos(\omega t)$, then the phase variation $\Delta \phi(t)$, which is
detectable by interferometry, amounts to 
$\Delta\phi(t)=\frac{4\pi}{\lambda} \, X \, \cos(\omega t)$
for a simple reflection geometry. Sub-picometer phase sensitivity is not needed
to detect the predicted hundreds-of-nanometers signals; rather, the
interferometer supplies large dynamic range and precise phase-coherent tracking
while the uncertainty is dominated by geometry and source characterization.

\subsection{Scope and Organization of the Paper}

The scope of the paper is to describe a  
general layout for an optimized space-based approach
to the measurement of $G$, not to offer a discussion
of all questions related to detailed engineering,
verification, and qualification. Also, a preliminary mission
design is beyond the scope of the current study.
We only discuss the measurement principle, a general layout 
with three test and two sources masses in a configuration that
allows for the mitigation of a number of systematic effects,
together with a concept study that includes the 
signal magnitude and its scaling, and selected conceivable 
mission parameters such as a suitable location in geostationary 
orbit. Further, more detailed engineering tasks and verification 
questions are relegated to future investigations.

The paper is organized as follows.
In Sec.~\ref{sec2}, we describe the proposed 
instrument architecture, which involves 
three test and two source masses, and two 
observation channels, the DC and the LC channel.
Symmetries of the apparatus and corresponding 
selection rules are discussed in Sec.~\ref{sec3}.
In Sec.~\ref{sec4}, we describe, in some detail,
the conversion of the covariances of the observations
into covariances of the underlying parameters of the
experiment; the latter include the gravitational constant.
The method of covariances is, in principle, well known, but
we believe that it does not hurt at all to
provide an illustrative numerical example, inspired
by the proposed experimental geometry, in Sec.~\ref{sec5}.
In Sec.~\ref{sec6F}, the consistency of the
results for $G$ from the DC and LC channels is
being analyzed, a task which requires the use
of cross covariances between the DC-only and LC-only channels. 
Mission parameters are discussed in Sec.~\ref{sec7},
while conclusions are reserved for Sec.~\ref{sec8}.

\begin{figure*}[t!]
\begin{center}
\begin{minipage}{0.99\linewidth}
\begin{center}
\includegraphics[width=0.8\linewidth]{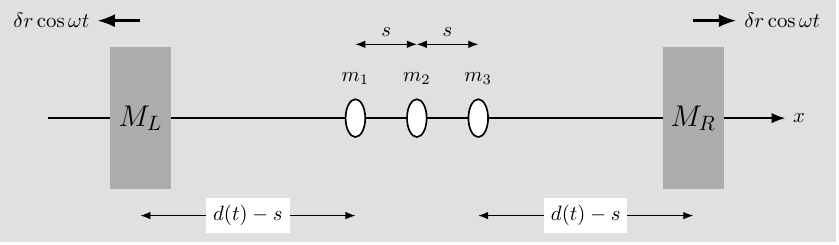}
\end{center}
\caption{\label{fig1}In the idealized symmetric geometry
of the space-based apparatus, the test masses are 
free during data taking. Capacitive electrodes and 
release mechanisms are omitted from the figure for clarity.
The left and right source masses are labeled $M_L$ and $M_R$,
while the left and right test masses are $m_1$ and $m_3$, with the 
central reference test mass being labeled $m_2$.
In the LC channel, the source masses undergo oscillatory
motion with amplitude $\delta r$ and angular 
frequency $\omega$, with their displacement begin denoted as $d(t)$.}
\end{minipage}
\end{center}
\end{figure*}

%
% Instrument Architecture
%
\section{Instrument Architecture}
\label{sec2}

%
% Test-Mass Coordinates
%
\subsection{Test-Mass Coordinates}
\label{sec2A}

We consider the schematic arrangement in Fig.~\ref{fig1}.
Let $\mathcal X_i(t)$, $i=1,2,3$, denote the signed
Cartesian positions ($\mathcal X_i(t)$ can
be negative) of the three test masses $m_i$ in an inertial coordinate system.
While the presence of three test masses 
(rather than two) may increase the 
complexity of the apparatus as compared to the 
Cavendish experiment~\cite{Cavendish1798}, the added test
mass offers the possibility of introducing a null
coordinate which can be used to eliminate 
a number of potential systematic effects from the 
experiment, as will be explained in further detail in the 
following. The indices $i=1,2,3$ refer to the left, central, and
right test masses,
\begin{equation}
x_1=x_L,
\qquad
x_2=x_C,
\qquad
x_3=x_R.
\end{equation}
The nominal positions are denoted by a superscript $(0)$,
and hence,
\begin{equation}
x_1^{(0)}=-s,
\qquad
x_2^{(0)}=0,
\qquad
x_3^{(0)}=+s \,,
\end{equation}
where $s>0$ is the nominal arm length.
The actual positions are written as
\begin{equation}
x_i(t) = x_i^{(0)} + \xi_i(t),
\end{equation}
where $\xi_i(t)$
denotes the time-dependent departure of test mass \(i\) from its
nominal position. The two (positive) arm lengths are
\begin{equation}
\ell_L(t)=x_2(t)-x_1(t),
\qquad
\ell_R(t)=x_3(t)-x_2(t).
\end{equation}
In the nominal configuration,
$\ell_L^{(0)} = \ell_R^{(0)} = s$.
The (time-dependent) signal coordinate is defined by
\begin{equation}
y_{\rm S}(t) = x_3(t)-x_1(t) = \ell_L(t)+\ell_R(t).
\end{equation}
Its nominal value is $y_{\rm S}^{(0)}=2s$.
The null coordinate is
\begin{equation}
y_{\rm N}(t) = x_1(t)+x_3(t)-2x_2(t) = \ell_R(t)-\ell_L(t).
\end{equation}
Its nominal value vanishes, $y_{\rm N}^{(0)}=0$.
The quantities \(y_{\rm S}(t)\) and \(y_{\rm N}(t)\) are instantaneous
coordinate combinations;  they should not be confused with the
processed observables introduced below.

\begin{figure*}[ht]
\begin{center}
\begin{minipage}{0.99\linewidth}
\begin{center}
\includegraphics[width=0.8\linewidth]{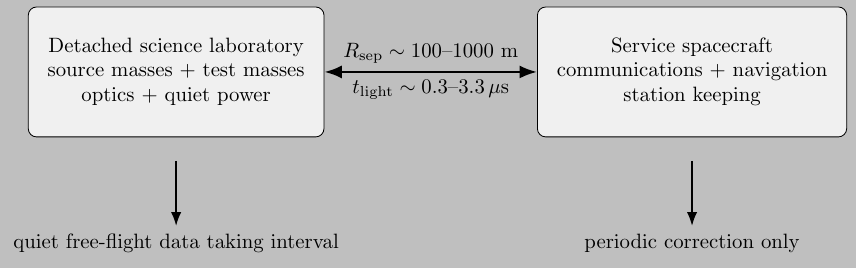}
\end{center}
\caption{\label{fig2}In the proposed
detached-laboratory architecture,
the service spacecraft is removed from the data taking
apparatus because the former can represent 
a nearby gravitational, mechanical, and thermal perturbation 
for the latter, during data taking.}
\end{minipage}
\end{center}
\end{figure*}

\subsection{DC Channel: Trajectory}
\label{sec2C}

In the DC channel,
the source masses are fixed at $\pm d$.
For a test mass at $x$, the (axial) acceleration is
\begin{equation}
\label{eq:axexact}
a(x;d) = GM\left[\frac{1}{(d-x)^2}-\frac{1}{(d+x)^2}\right] 
=\frac{4 \, d \, G \, M \; x}{(d^2-x^2)^2}.
\end{equation}
The center mass has $a(0;d)=0$. For the outer test masses at $\pm s$,
when released from their nominal positions,
the differential acceleration is therefore
\begin{equation}
A(d)\equiv a(+s;d)-a(-s;d)
= \frac{8 \, d \, GM \, s}{(d^2-s^2)^2}.
\label{eq:Ad}
\end{equation}
For $s/d\ll1$,
\begin{equation}
A(d)=\frac{8GMs}{d^3} \, 
\left[1+2\frac{s^2}{d^2}+3\frac{s^4}{d^4}+\cdots\right].
\label{eq:ADCsmall}
\end{equation}
The leading term is the source-generated gravity gradient multiplied by the
test-mass baseline $2s$.

With $d=d_0$, the ideal DC differential acceleration is
\begin{equation}
A_0=\frac{8GMd_0s}{(d_0^2-s^2)^2}.
\label{eq:A0}
\end{equation}
If the test masses begin with differential position $D_0$ and velocity $V_0$, a
short free-flight segment has approximately
\begin{equation}
y_{\rm S}(t) = X_0 + V_0 \, t + \frac12 A_0 \, t^2+\cdots,
\label{eq:parabola}
\end{equation}
where $X_0 = y_{\rm S}(t=0)$ and
omitted terms include spatial variation of the source field, orbital
tides, and nuisance accelerations. 
A static measurement can be very strong statistically because displacement
grows as $t^2$. It is also vulnerable to low-frequency drifts and degeneracies
with slowly varying spacecraft gravity and orbital tides. These limitations
motivate the modulated channel.

%
% DC Channel: Observable
%
\subsection{DC Channel: Observable}
\label{sec2D}

The relevant observable in the DC channel is the best
estimate $\widehat A_0$ for the acceleration due to gravity.
The DC channel is obtained from the same time series 
through a dynamical fit. A local model may be written 
\begin{equation}
 \ddot y_{\rm S}(t) = A_0 + K_D \, y_{\rm S}(t) +
\Gamma_{\oplus}(t) \, y_{\rm S}(t) + a_{\rm par}(t),
\end{equation}
where $K_D$ is a coefficient that 
describes local stiffness, $\Gamma_{\oplus}$ is the projected
Earth-tide tensor (here, ``tide'' is to be 
understood in the sense of an astrophysical 
jargon, where it refers to the
variation of the Earth's gravitational
field over the dimensions of the apparatus,
not in the sense of ocean ``tides'').
Furthermore, $a_{\rm par}$ collects parasitic differential
accelerations. 

For long arcs, it can be advantageous to 
sample the trajectory of the test masses
at defined sampling times, and to calculate
approximations for the integrals of $y_{\rm S}(t) \, t^n$ 
with $n=0,1,2,\dots$. It is then possible to 
infer the parameters $X_0$, $V_0$ and $A_0$ 
by solving a linear system of equations.
For example, if we neglect 
terms of higher order than $t^2$ in 
Eq.~\eqref{eq:parabola}, and set
\begin{equation}
J_k = \int_0^{t_0} y_{\rm S}(t) \, \left( \frac{t}{t_0} \right)^n \, \dd t 
\,, \qquad n = 0,1,2 \,,
\end{equation}
where $t_0$ marks the end of the sampling interval,
then
\begin{equation}
\left( \begin{array}{c}
X_0 \, t_0 \\ V_0 \, (t_0)^2 \\ A_0 \, (t_0)^3
\end{array} \right) =
\left( \begin{array}{ccc}
9 & -36 & 30 \\
-36 & 192 & -180 \\
60 & -360 & 360 \\
\end{array} \right) \,
\left( \begin{array}{c}
J_0 \\ J_1 \\ J_2 
\end{array} \right) \,.
\end{equation}
Further details regarding the 
optimum extraction of the observable
\begin{equation}
y_{\rm DC} = \widehat A_0
\label{yDC}
\end{equation}
are beyond the scope of this paper. 
For clarification, the symbols $y_{\rm DC}$ denotes
an entry in the vector of observables
that enter the final evaluation of the 
experiments using the method 
of covariances [see also Eq.~\eqref{full_observables}]; 
hence, $y_{\rm DC}$ has a different physical 
dimension (that of acceleration) as compared to the 
signal coordinate $y_{\rm S}$. One observes that 
the integrals $J_n$ can be approximated as follows,
\begin{equation}
J_n \simeq \sum_{k = 0}^{N-1} w_k \,
y_{\rm S}(t_k) \, \left( \frac{t_k}{t_0} \right)^n
\,, \quad n = 0,1,2 \,,
\end{equation}
where the weights $w_k$ sample the integration
interval. A possible choice for $N$ integration nodes is
the obvious equidistant one, namely,
$w_k = \Delta t$ with $t_k = k \, \Delta t$
for $k = 0, \dots, N-1$, and $N \, \Delta t = t_0$.

\subsection{LC Channel: Trajectory}
\label{sec2E}

For the LC channel, we assume an oscillatory motion of the source masses.
Let
\begin{equation}
d(t) = d_0 + q(t) \,,
\qquad q(t) = \delta r \cos(\omega t + \phi) \,.
\end{equation}
Expanding $A(d(t))$ about the nominal point
$d(t) = d_0$ gives the time-dependent 
acceleration $A(t)$ for the signal coordinate
$y_{\rm S}(t)$,
\begin{equation}
A(t)=A_0 + A'_0 \, q(t) + \frac12 \, A''_0 \, q^2(t) + \dots,
\end{equation}
where derivatives are evaluated at $d_0$. 
The acceleration amplitude for the 
acceleration of the signal coordinate $y_{\rm S}(t)$ is, 
to first order in $\delta r$,
given by the expression
\begin{equation}
A'_0 \, q(t) = 
A_\omega \, \cos(\omega t + \phi) \,,
\end{equation}
where
\begin{equation}
A_\omega = 
%%% A'_0 \, \delta r =
-8 \, G \, M \, \frac{s \, (3d_0^2+s^2)}{(d_0^2-s^2)^3} \, \delta r \, .
\label{eq:Aomega}
\end{equation}
If the nominal distance between the 
test masses is small against the 
nominal separation of the source masses ($s\ll d_0$),
then
\begin{equation}
A_\omega\simeq - 24 \, G M \, \frac{s}{d_0^4} \, \delta r \,.
\label{eq:Aomegasmall}
\end{equation}
The sign indicates that outward 
motion reduces the differential attraction.
For undamped motion (e.g., motion induced 
by precise actuators), the displacement amplitude is 
\begin{equation}
D_\omega \simeq -\frac{A_\omega}{\omega^2} \,.
\label{eq:Domega}
\end{equation}
If the motion of the source masses is induced 
via a spring, then damping will occur.
The solution of the equation then needs to be 
modified to account for the damping parameters,
which is a straightforward calculation 
that can easily be accomplished on the basis 
of the Green's function of the damped 
harmonic oscillator~(see Chap.~2 of Ref.~\cite{Je2017book}).

Dividing the result given in Eq.~\eqref{eq:Aomega} by 
the one from Eq.~\eqref{eq:A0} yields the first-order ratio
\begin{equation}
\frac{A_\omega}{A_0} = -2\delta r\frac{3d_0^2+s^2}{d_0(d_0^2-s^2)}
\simeq \frac{A_\omega}{A_0}\simeq-3\frac{\delta r}{d_0}.
\label{eq:ratio}
\end{equation}
The ratio is independent of $G$ and $M$. It is therefore an unusually valuable
geometry and modulation calibration observable. Agreement between the measured
ratio and the source-position metrology tests the point-source model and the
relative calibration of the DC and LC analyses.

A remark should be added concerning higher harmonics.
Namely, since $q^2=(\delta r)^2[1+\sin(2\omega t)]/2$, the second harmonic 
has the amplitude
\begin{equation}
A_{2\omega} = \frac14 \, A''_0 \, (\delta r)^2 +
{\mathcal O}(\delta r^4).
\end{equation}
Odd and even harmonics encode different symmetry information. The fundamental
is linear in modulation amplitude; the second harmonic is quadratic.
While we do not dwell on this aspect 
in any further detail, one should be aware that
higher harmonics could be retained rather than filtered away.
This would enable one to carry out further checks,
e.g., by distinguishing the gravitational effects
(contributions to $A''_0$ proportional to $GM$)
from actuator-correlated disturbances which could
manifest themselves in 
spurious contributions to $A''_0$ that scale differently.

\subsection{LC Channel: Observable}

The solution to the differential equation
\begin{equation}
\ddot y_{\rm S}(t) = A_\omega \, \cos(\omega t + \phi) \,,
\end{equation}
with the initial conditions $y_{\rm S}(t=0) = 0$ and
$\dot y_{\rm S}(t=0) = 0$, is
\begin{equation}
y_{\rm S}(t) = - \frac{A_\omega}{\omega^2} 
\left[ \cos(\omega t + \phi) - \cos(\phi) +
\omega \, t \, \sin(\phi) \right] \,.
\end{equation}
For nonzero slippage phase $\phi$, 
there is a small residual linear term proportional to $t$
remaining in the trajectory of the signal coordinate.
For $T = 2 \pi/\omega$, and $n=0,1,2,\dots$, one 
can easily show that
\begin{subequations}
\begin{align}
C_n =& \; \int_0^{n T} \, y_{\rm S}(t) \, \cos(\omega t) \, \dd t =
-n \pi \frac{A_\omega}{\omega^3} \, \cos(\phi) \,,
\\
S_n =& \; \int_0^{n T} \, y_{\rm S}(t) \, \sin(\omega t) \, \dd t = 
3 n \pi \frac{A_\omega}{\omega^3} \, \sin(\phi) \,,
\end{align}
\end{subequations}
The observable
\begin{equation}
y_{\rm LC} = \widehat A_\omega
\label{yLC}
\end{equation}
can be obtained from
\begin{equation}
A_\omega = \frac{\omega^3}{3 \pi n} \,
\sqrt{ 9 \, (C_n)^2 + (S_n)^2 } \,,
\end{equation}
while $S_n/C_n = -3 \tan(\phi)$.

For a sampled differential trajectory $y_{\rm S}(t_k)$, 
one can obtain weighted quadratures
for $C_n$ and $S_n$ using the formulas,
\begin{subequations}
\begin{align}
C_n \simeq & \; \sum_k w_k \, y_{\rm S}(t_k) \, \cos(\omega t_k) \,,
\\
S_n \simeq & \; \sum_k w_k \, y_{\rm S}(t_k) \, \sin(\omega t_k) \,.
\end{align}
\end{subequations}
A possible choice is $w_k = \Delta t$ with $t_k = k \, \Delta t$
for $k = 0, \dots, N-1$, and $N \, \Delta t = n T = 2 \pi n/\omega$.
This choice is fortunate because the quadrature error
for the cosine and sine functions vanishes,
\begin{equation}
\frac{1}{N} \sum_{k=0}^{N-1} \cos^2(\omega t_k) 
= \frac{1}{N} \sum_{k=0}^{N-1} \sin^2(\omega t_k) 
= \frac{1}{2}.
\end{equation}
In other words, equal 
integration weights are particularly useful because they preserve the exact
discrete orthogonality of the sine and cosine templates over complete
cycles. The sampled cosine and sin functions are then discretely orthogonal:
\begin{equation}
\sum_{k=0}^{N-1}
\cos(\omega t_k)\sin(\omega t_k) = 0 \,.
\end{equation}

The DC channel contains high signal and strong cumulative displacement, but it
is correlated with slowly varying backgrounds. The lock-in channel is
spectrally isolated and supplies phase and harmonic diagnostics, but it depends
more strongly on modulation geometry and source-motion metrology. A
convincing result requires that the values for $G$ obtained 
from both channels be mutually consistent,
$G_{\DC}=G_{\rm LC}$.  A comprehensive discussion of the 
consistency requires an analysis based
on cross covariances (see also Sec.~\ref{sec6F}).

%
% Symmetries and Parameter Space
%
\section{Symmetries and Parameter Space}
\label{sec3}

%
% Reflection Symmetry
%
\subsection{Reflection Symmetry}
\label{sec3A}

The previous section introduced the geometrical configuration of the experiment
together with the signal and null coordinates.  In order to classify the
observables and nuisance parameters relevant to the experiment, it is
advantageous to exploit the approximate reflection symmetry of the apparatus.
As will become apparent below, this symmetry determines much of the structure
of the Jacobian matrix and consequently of the covariance matrix obtained from
the generalized least-squares adjustment.  Thus, the purpose of this section is
to carefully formulate the reflection properties of the test-mass and
source-mass coordinates in the proposed space-based measurement of Newton's
gravitational constant $G$.

The central point is that a reflection about the geometrical midpoint of the
apparatus consists of two operations, namely, the inversion of the signed
Cartesian coordinate, $x \longmapsto -x$, and the interchange of the physical
left and right locations.  The left and right labels therefore cannot be kept
attached to the same physical objects while the Cartesian coordinates are
merely multiplied by $-1$.  Instead, the reflected right-hand position
becomes the new left-hand position, and the reflected left-hand position
becomes the new right-hand position.  Once this distinction is made, the
transformation properties of the signal coordinate, the null coordinate, and
the source-center displacement parameters follow unambiguously.

After careful consideration, it becomes
apparent that the reflection transformation $\mathcal R$ 
of the ordered coordinate triple can be described 
as follows,
\begin{equation}
\mathcal R:
\quad
(x_L,x_C,x_R)
\longmapsto
(-x_R,-x_C,-x_L) \,,
\end{equation}
or equivalently, with primed and unprimed coordinates,
\begin{equation}
x_L'=-x_R, \quad x_C'=-x_C, \quad x_R'=-x_L.
\end{equation}

%
% Reflection and Arm Length
%
\subsection{Reflection and Arm Lengths}
\label{sec3B}

Let the signed Cartesian coordinates (they can 
be negative) of the three test masses be
$x_L$, $x_C$, and $x_R$, with the ordering
$x_L < x_C < x_R$.
In the ideal reference configuration, one may take
\begin{equation}
x_L^{(0)}=-s,
\qquad
x_C^{(0)}=0,
\qquad
x_R^{(0)}=+s,
\end{equation}
where \(s>0\).
Let \(\mathcal R\) denote reflection about the geometrical midpoint
\(x=0\).
A point at coordinate \(x\) is mapped to the coordinate \(-x\).
However, after reflection, the point originally on the right lies on
the left, and the point originally on the left lies on the right.
The primes denote coordinates in the reflected configuration.
It is useful to verify that the ordering is preserved.
Namely, with $x_L<x_C<x_R$, one obtains
$-x_R<-x_C<-x_L$, so that $x_L'<x_C'<x_R'$.
Thus, the primed labels again refer to the left, center, and right
locations of the reflected apparatus.

We define the positive-definite arm lengths by
\begin{equation}
\ell_L=x_C-x_L, \qquad \ell_R=x_R-x_C \, .
\end{equation}
Because $x_L<x_C<x_R$,
both quantities are positive,
$\ell_L>0$, and $\ell_R>0$.
In the symmetric reference configuration,
$\ell_L^{(0)} = x_C^{(0)}-x_L^{(0)} = 0-(-s) = s$,
and $\ell_R^{(0)} = x_R^{(0)}-x_C^{(0)} = s-0 = s$.
Hence,
\begin{equation}
\ell_L^{(0)}=\ell_R^{(0)}=s \,.
\end{equation}

Let us exercise our understanding of 
the symmetry transformation
by considering the transformation of the left arm.
In the reflected configuration, one has
\begin{equation}
\ell_L' = x_C' - x_L' = (-x_C) - (-x_R) = \ell_R \, .
\end{equation}
Therefore, we have $\ell_L' = \ell_R$.
Let us also consider the transformation of the right arm.
Similarly, one finds
\begin{equation}
\ell_R' = x_R'-x_C' =
(-x_L) - ( -x_C ) = \ell_L.
\end{equation}
Combining the two results gives
\begin{equation}
\mathcal R: \quad 
\ell_L \longmapsto \ell_R \, \qquad
\ell_R \longmapsto \ell_L \, .
\end{equation}
Thus, reflection simply exchanges the two positive arm lengths.

It is our task now to evaluate the symmetry properties
of the signal and null coordinates under the reflection,
\begin{equation}
y_{\rm S} =x_R-x_L,
\quad
y_{\rm N} =x_L+x_R-2x_C \,.
\end{equation}
These definitions have a direct interpretation in terms of the arm
lengths. One finds that
\begin{equation}
y_{\rm S} = \ell_L+\ell_R \,,
\quad
y_{\rm N} = \ell_R-\ell_L \,.
\end{equation}
The null coordinate measures the imbalance between the right and left
arms.

%
% Signal and Null Coordinates
%
\subsection{Signal and Null Coordinates}
\label{sec3C}

From the arm-length representation, we infer that 
\begin{equation}
\label{ySreflection}
y_{\rm S}' = \ell_L'+\ell_R' = \ell_R+\ell_L = y_{\rm S} \,.
\end{equation}
Therefore, $y_{\rm S}$ belongs to the reflection-even sector.
For the null coordinate, we have the
transformation property
\begin{equation}
\label{yNreflection}
y_{\rm N}' = \ell_R'-\ell_L' =
-(\ell_R-\ell_L) = -y_{\rm N} \,.
\end{equation}
Therefore, $y_{\rm N}$ belongs to the reflection-odd sector.

%
% Source-Mass Center Coordinates
%
\subsection{Source-Mass Center Coordinates}
\label{sec3D}

Let the nominal source-mass centers be located at
$X_L = -d_0$ and $X_R = +d_0$.
Let the actual effective gravitational-center coordinates be
\begin{equation}
X_L=-R+\delta_L, \qquad
X_R=+R+\delta_R \,.
\end{equation}

Here, $\delta_L$ and $\delta_R$ are signed Cartesian displacements
measured in the same global \(x\) direction.
Thus, a positive value of either \(\delta_L\) or \(\delta_R\) means a
displacement toward increasing \(x\).
Under reflection, the original right source becomes the new left
source.  Therefore,
\begin{equation}
-R+\delta_L' \equiv X'_L = -X_R = -( R+\delta_R ) = -R-\delta_R .
\end{equation}
Hence, $-R+\delta_L' = -R-\delta_R$ and thus
\begin{equation}
\delta_L'=-\delta_R \,.
\end{equation}
Similarly, the original left source becomes the new right source:
\begin{equation}
R+\delta_R' \equiv X'_R = -X_L = -(-R+\delta_L)
= R-\delta_L.
\end{equation}
Hence, $R+\delta_R' = R-\delta_L$ and thus
\begin{equation}
\delta_R'=-\delta_L \,.
\end{equation}
Combining the two transformations,
one arrives at the conclusion that both
$\delta_L$ as well as $\delta_R$ change sign
under $\mathcal R$:
\begin{equation}
\mathcal R: \qquad
\delta_L\longmapsto-\delta_R, \qquad \delta_R\longmapsto-\delta_L.
\end{equation}
The minus signs are essential because \(\delta_L\) and \(\delta_R\)
are signed Cartesian displacements.

One can now define symmetric and antisymmetric source parameters
as follows,
\begin{equation}
\label{deltaPMdef}
\delta_+ = \frac{\delta_L+\delta_R}{2},
\qquad
\delta_- = \frac{\delta_R-\delta_L}{2}.
\end{equation}
One now verifies easily that, under reflection,
\begin{equation}
\mathcal R:
\qquad
\delta_+ \longmapsto -\delta_+ \,,
\qquad
\delta_- \longmapsto \delta_-.
\end{equation}
Thus \(\delta_-\) is reflection even.
More explicitly, one has the following
reflection-odd relations
\begin{equation}
\mathcal R:
\qquad
G \longmapsto G,
\qquad
y_{\rm S} \longmapsto y_{\rm S},
\qquad
\delta_- \longmapsto \delta_- \,,
\end{equation}
and the following reflection-even ones,
\begin{equation}
\mathcal R:
\qquad
y_{\rm N} \longmapsto -y_{\rm N},
\qquad
\delta_+ \longmapsto -\delta_+.
\end{equation}

%
% Selection Rules for the Jacobian
%
\subsection{Selection Rules for the Jacobian}
\label{sec3E}

Let the observables [we have
introduced two of these in 
Eqs.~\eqref{yDC} and~\eqref{yLC}]
be functions $f_i$ of the model parameters,
which are summarized in a vector $\bm{p}$,
\begin{equation}
\label{JACdef}
y_i=f_i(\bm p), \qquad A_{ij} = \left.
\frac{\partial y_i}{\partial p_j}
\right|_{\bm p=\bm p_0},
\end{equation}
where $\bm{A}$ is the Jacobian matrix
(we here denote both vectors and matrices in boldface)
and \(\bm p_0\) denotes the symmetric reference configuration.
We can anticipate here that the following
Jacobian, which relates the observables and 
the model parameters, will be of utmost
importance in our discussions. At the symmetric reference point, the 
geometry of the apparatus respects the
reflection operation \(\mathcal R\).  

A first-order variation can couple
an observable and a parameter only if they possess the same reflection
parity. Thus, even observables couple linearly to even parameters,
while odd observables couple linearly to odd parameters.
Linear couplings between quantities of opposite reflection parity
vanish at the symmetric reference point.

Let now $y_{\rm S}$ denote a generic 
observable derived from the signal coordinate,
such as $y_{\rm DC}$ or $y_{\rm LC}$.
Because $y_{\rm S}$ and $\delta_-$ are even,
the following partial derivatives
\begin{equation}
\label{ySNpartials}
\left.
\frac{\partial y_{\rm S}}{\partial\delta_-}
\right|_0 \neq 0 \,, \qquad
\left.  \frac{\partial y_{\rm N}}{\partial\delta_+}
\right|_0 \neq 0 \,,
\end{equation}
are symmetry allowed.
We can also give examples for forbidden crossed couplings.
Since $y_{\rm S}$ is even while $\delta_+$ is odd,
we have, at the symmetric reference configuration,
denoted here by a subscript zero,
\begin{equation}
\left.
\frac{\partial y_S}{\partial\delta_+}
\right|_0 =
\left.
\frac{\partial y_{\rm N}}{\partial\delta_-}
\right|_0 = 0 \,.
\end{equation}
These are first-order reflection selection rules.
It is extremely instructive to study the 
partial derivative of the null coordinate with respect to $G$.
On the one hand, $G$ (a fundamental constant) 
is certainly invariant under reflection,
$\mathcal R: G \longmapsto G$.
On the other hand, the null coordinate 
\(y_{\rm N}\) is known to be reflection odd,
$\mathcal R: y_{\rm N} \longmapsto -y_{\rm N}$.
A variation of the even parameter \(G\) cannot generate an odd response
in an exactly symmetric reference configuration.  Consequently,
in view of the symmetry properties under reflection,
\begin{equation}
\label{partial}
\left.
\frac{\partial y_{\rm N}}{\partial G}
\right|_0 = 0.
\end{equation}
This result can also be seen directly from the Newtonian accelerations.
Let $g_L$, $g_C$, and $g_R$
denote the source-generated accelerations of the three test masses.
The second derivative of the null coordinate is
\begin{equation}
\ddot y_{\rm N} =
\ddot x_L+\ddot x_R-2\ddot x_C
= G\left( h_L+ h_R-2 h_C \right).
\end{equation}
At the symmetric reference point,
$h_L = -h_R$, $h_C=0$, and thus $\ddot y_{\rm N}=0$,
for every value of \(G\). The relation~\eqref{partial} is thus
confirmed. Thus, the vanishing derivative 
given in Eq.~\eqref{partial} may be understood either as a
reflection-parity selection rule or as an explicit cancellation of the
Newtonian accelerations.

One caveat is of utmost importance.
Namely, the actual apparatus will not be perfectly reflection symmetric.
Small differences may arise from
$M_L\neq M_R$, $Q_L\neq Q_R$, and $\delta_+\neq0$,
misalignments, thermal gradients, spacecraft gravity, or other
imperfections.
Let \(\epsilon_{\rm odd}\) denote a generic reflection-odd imperfection,
an example being the functional form $M_L = M_R \, (1 + \epsilon_{\rm odd})$.
Then, an otherwise forbidden coupling can appear at the 
order of the symmetry-breaking parameter $\epsilon_{\rm odd}$,
\begin{subequations}
\begin{align}
\frac{\partial y_{\rm N}}{\partial G}
=& \; \mathcal O(\epsilon_{\rm odd}) \,,
\\
\frac{\partial y_{\rm S}}{\partial\delta_+}
=& \; \mathcal O(\epsilon_{\rm odd}) \,,
\qquad
\frac{\partial y_{\rm N}}{\partial\delta_-}
= \mathcal O(\epsilon_{\rm odd}) \,.
\end{align}
\end{subequations}
The zeros in the ideal Jacobian should therefore be interpreted as
exact zeros at the symmetric reference point and as parametrically
small entries in the real apparatus.

\subsection{Geometric Selection Rules}
\label{sec3F}

{\em Underlying Rationale.---}In additional to the signal and null coordinates
$y_{\rm S}$ and $y_{\rm N}$,
and the source mass displacements $\delta_L$ and 
$\delta_R$, which were discussed in the last section, 
there are additional geometric factors 
whose symmetry properties under reflection
are of interest for our studies.
In the current section, we focus on the 
multipole moments of the source mass,
and on the gravity gradient (``tidal variation
of the gravitational field'') across the spaceship.
These parameters are, {\em a priori}, nuisance parameters.
However, in the context of the covariance analysis,
the introduction of these parameters helps in 
distinguishing the effects of the gravitational
constant from geometric and other disturbances.
Rather than trying to eliminate all systematics
from the experiment, we shall attempt to 
estimate many of them using a generalized least-squares
adjustment.

{\em Multipole Corrections.---}The source masses 
are not assumed to be perfect point masses.
Outside each source, with $r$ the distance
of the source from the observation point,
the gravitational potential may be written as
\begin{equation}
\Phi = - \frac{GM}{r}
- G \sum_{\ell=2}^{\infty}
\frac{1}{r^{\ell+1}}
\sum_{m=-\ell}^{+\ell}
Q_{\ell m}
Y_{\ell m}(\theta,\phi).
\end{equation}
For the present analysis it is sufficient to retain only the leading
quadrupole correction,
$Q_{20}$, which we assume to be oriented along the 
symmetry axis of the two near-perfect source mass spheres
as the crystals were grown in the gravitational field
of the Earth
The quadrupole field constitutes the dominant departure from spherical symmetry.
Higher multipoles may be incorporated straightforwardly if required.

{\em Quadrupole Moments.---}The two source masses possess, 
in general, different quadrupole
moments, $Q_{20}^{(L)}$ and $Q_{20}^{(R)}$.
Under the generalized reflection operation,
the left and right source masses are interchanged,
\begin{equation} 
{\mathcal R}: \;\; Q_{20}^{(L)} \longmapsto Q_{20}^{(R)} \,,
\quad
Q_{20}^{(R)} \longmapsto Q_{20}^{(L)} \,.
\end{equation} 
No additional minus sign appears;
the quadrupole tensor is a second-rank tensor, and therefore remains
unchanged under spatial inversion apart from the interchange of the two
source masses.

One may thus introduce the symmetric and antisymmetric combinations
\begin{equation}
\label{Q20PMdef}
Q_{20}^{(+)}
= \frac{ Q_{20}^{(L)} + Q_{20}^{(R)} }{2} \,,
\qquad
Q_{20}^{(-)}
= \frac{ Q_{20}^{(R)} - Q_{20}^{(L)} }{2} \,,
\end{equation}
with the transformation properties
\begin{equation}
{\mathcal R}: \;\; 
Q_{20}^{(+)} \longmapsto Q_{20}^{(+)},
\quad
Q_{20}^{(-)} \longmapsto -Q_{20}^{(-)}.
\end{equation}

{\em Spacecraft Gravity.---}Even in 
the absence of the movable source masses, the spacecraft itself
generates a gravitational field.
Furthermore, the gravitational field
of conceivable nearby satellites, and the tidal variations
of the gravitational field of the Earth itself,
generate additional field gradients which can be 
incorporated into our formalism.
Near the test masses, the residual field may be expanded as
\begin{equation}
g_x(x) = g_0 +
\Gamma_{xx} \, x + \frac12 \, H_{xxx} \, x^2 + \dots ,
\end{equation}
where $g_0$ is the field at the location
of the geometric center of the test mass, 
$x$ is a generic variable denoting a departure
from the reference point, and 
\begin{equation}
\label{GammaXXdef}
\Gamma_{xx} = \frac{\partial g_x}{\partial x}
\end{equation}
denotes the local gravity gradient.
The constant term $ g_0$
produces only a common acceleration of the test masses and therefore
does not contribute directly to the differential observable.
The gravity gradient, however, generates differential accelerations
between the test masses and consequently enters the measurement model
explicitly.

{\em Gravity-Gradient Parameters.---}The leading gradient
terms $\Gamma_{xx}^{(L)}$ and $\Gamma_{xx}^{(R)}$
are mapped as follows under reflection,
\begin{equation}
{\mathcal R}: \;\;
\Gamma_{xx}^{(L)} \longmapsto \Gamma_{xx}^{(R)},
\quad
\Gamma_{xx}^{(R)} \longmapsto \Gamma_{xx}^{(L)},
\end{equation}
The symmetric and antisymmetric combinations,

\begin{equation}
\label{GammaXXPMdef}
\Gamma_{xx}^{(+)} = \frac{ \Gamma_{xx}^{(L)} + \Gamma_{xx}^{(R)} }{2},
\qquad
\Gamma_{xx}^{(-)} =
\frac{ \Gamma_{xx}^{(R)} - \Gamma_{xx}^{(L)} }{2},
\end{equation}
transform as follows,
\begin{equation}
{\mathcal R}: \;\;
\Gamma_{xx}^{(+)} \longmapsto \Gamma_{xx}^{(+)} \,,
\quad
\Gamma_{xx}^{(-)} \longmapsto -\Gamma_{xx}^{(-)} \,.
\end{equation}

In the following, we shall also consider a background coordinate
$y_{\rm BG}$ (measured with the source
masses removed or placed far away from the test masses)
and a rotated coordinate $y_{\rm rot}$
(measured with the source masses rotated, mainly 
designed to be sensitive to multipole contributions
to the source-mass distribution).
We assume that the rotated and background coordinates
$y_{\rm rot}$ and $y_{\rm BG}$ are defined 
in analogy to the signal coordinate $y_{\rm S}$.
Hence, 
\begin{equation}
{\mathcal R}: \;\;
y_{\rm rot} \longmapsto y_{\rm rot},
\quad
y_{\rm BG} \longmapsto y_{\rm BG} \,.
\end{equation}

Let us also introduce a parameter $\eta_{\rm th}$,
$\eta_{\rm th}$, which describes a thermal correction 
that leads to a tiny, temperature-dependent acceleration
according to 
\begin{equation}
\label{etaTHdef}
a_{\rm th} = \eta_{\rm th}\, \Phi_{\rm th} \,.
\end{equation}
While $\eta_{\rm th}$ possesses no intrinsic reflection parity,
One decomposes the field into 
its symmetric and antisymmetric components,
\begin{equation}
\Phi_{\rm th} = \Phi^{(+)} + \Phi^{(-)},
\end{equation}
so that, under reflection,
\begin{equation}
\mathcal R: \;\;
\Phi^{(+)} \longmapsto \Phi^{(+)},
\quad
\Phi^{(-)} \longmapsto - \Phi^{(-)}.
\end{equation}

%
% Observables and Parameters
%
\section{Observables and Parameters}
\label{sec4}

%
% Extended List of Observables
%
\subsection{Extended List of Observables}
\label{sec4A}

A central paradigm of our studies is to 
apply the method of a generalized least-squares
adjustment to an over-determined
system, where the number of observables 
is greater than the number of parameters. We write the 
full vector $\mathbm y$
of experimentally determined observables is written as
\begin{equation}
\label{full_observables}
\mathbm y = \left( \mathbm y_{\rm DC}, \mathbm y_{\rm LC}, \mathbm y_{\rm N}, 
\mathbm y^{{\rm rot}}_{\rm DC}, \mathbm y^{{\rm rot}}_{\rm LC}, 
\mathbm y^{{\rm rot}}_{\rm N}, y_{\rm BG}, y^{{\rm rot}}_{\rm BG} \right)^T .
\end{equation}
The components of $\mathbm y$ have been partially 
discussed in the preceding sections of this
article, but deserve
a more detailed discussion here.
A remark is in order: The list of observables is 
by no means exhaustive and could be supplemented
by further configurations, for example,
different orientations of the main laboratory with 
respect to the source masses, uncovering higher
multipole moments. However, we advocate the above vector given in
Eq.~\eqref{full_observables} as a possible first 
{\em ansatz} worthy of a consideration.

Now, let us discuss the components of $\mathbm y$.
The first component, in itself,
constitutes a vector 
$\mathbm y_{\rm DC}$,
with $K$ components for $K$ source-mass separations.
Each element of $\mathbm y_{\rm DC}$
contains the (integrated) differential accelerations
obtained from the simultaneous free release
of the three test masses in the static gravitational field of the
source masses (DC channel). 
We have described a possible extraction of the
acceleration signal in Sec.~\ref{sec2C}.
The observable could otherwise be obtained from the fitted quadratic
curvature of the signal coordinate.
Under rotation of the source masses,
one obtains the observables 
recorded in the vector $\mathbm y^{\rm rot}_{\rm DC}$.

The vector components of $\mathbm y_{\rm LC}$ are the analogs of 
$\mathbm y_{\rm DC}$, obtained from the lock-in channel.
As described in Secs.~\ref{sec2D} and~\ref{sec2E},
the acceleration amplitudes are obtained 
from the oscillatory (lock-in) signal generated by 
an induced, oscillatory motion of the source masses.
Under rotation of the source masses,
one obtains $\mathbm y^{\rm rot}_{\rm LC}$.

The null coordinates contained in $\mathbm y_{\rm N}$
(again, $K$ components for $K$ source-mass separations)
are used primarily for monitoring common-mode disturbances.
Under rotation of the source masses,
one obtains $\mathbm y^{\rm rot}_{\rm N}$.
As is evident from Eqs.~\eqref{ySreflection},~\eqref{yNreflection}
and~\eqref{ySNpartials}, the null coordinate,
which, for an ideal apparatus, should remain zero,
probes a different symmetry sector 
of the apparatus as compared to the signal 
coordinates and thus constitutes a sensitive sensor
for imperfections of the experimental setting. 

Background observables contained in the 
vector $\mathbm y_{\rm BG}$ are
obtained with the source masses removed from
their science configuration (or parked sufficiently far away).  This
observable is primarily sensitive to the residual spacecraft gravity
field and other static environmental perturbations.
Under rotation of the source masses,
one obtains $y^{\rm rot}_{\rm BG}$.

For $K$ distinct source-mass separations,
one obtains $6K + 2$ observables, which, for, say, $K = 3$,
amounts to 20 observables. This number exceeds the 
number of parameters in the vector of model parameters
[see Eq.~\eqref{full_parameters}],
leading to an overdetermined system (as it should).

%
% Extended List of Parameters
%
\subsection{Extended List of Parameters}
\label{sec4B}

We propose to formulate the complete estimation problem in terms of
a set of parameters similar to the following 
vector (which has $9$ components),
\begin{equation}
\label{full_parameters}
\mathbm p = \left( G, M_L, M_R, \delta_+, \delta_-, Q_{20}^{(+)},
Q_{20}^{(-)}, \Gamma_{xx}^{(+)}, \eta_{\rm th} \right)^T,
\end{equation}
where $G$ denotes Newton's gravitational constant,
$M_L$ and $M_R$ are the calibrated source masses,
and $\delta_+$ and $\delta_-$ describe the common and differential
effective gravitational centers,
as defined in Eq.~\eqref{deltaPMdef}.
The quadrupole components 
$Q_{20}^{(\pm)}$ of the source masses are 
defined in Eq.~\eqref{Q20PMdef}.
denote the symmetric and antisymmetric leading quadrupole
moments of the two source masses;

For the symmetric component of the 
gravity gradient $\Gamma_{xx}^{(+)}$, 
we refer to Eqs.~\eqref{GammaXXdef} and~\eqref{GammaXXPMdef}.
The parameter $\eta_{\rm th}$ describes 
a temperature-dependent residual contribution
to the acceleration, in accordance with Eq.~\eqref{etaTHdef}.

The elements of the parameter vector $\mathbm p$
are estimated simultaneously from the complete set of experimental
observables. They can be classified 
into three conceptually different classes of
parameters entering the observation model.
Although all of them appear on an equal footing in the parameter vector
$\mathbm p$, their physical interpretation and symmetry properties are quite
different.

%%%%%%%%%%%%%%%%%%%%%%%%%%%%%%%%%%%%%%%%%%%%%%%%%%%%%%%%%%%%%%%%%%%%%%%%%%

The first class (geometric parameters)
consist of parameters describing the geometry of the
apparatus itself, namely,
$\delta_\pm$, $Q_{20}^{(\pm)}$, and $\Gamma_{xx}^{(+)}$.
These quantities possess intrinsic transformation properties under the
generalized left--right reflection operation.
For example, $\delta_-$, $Q_{20}^{(+)}$, and
$\Gamma_{xx}^{(+)}$ are reflection even,
whereas $\delta_+$ and $Q_{20}^{(-)}$ are reflection odd.
The generalized reflection symmetry therefore acts directly on these
parameters.

The second class contains the physical constants and source-mass
parameters, $G$, $M_L$, and $M_R$.
Newton's gravitational constant is a true scalar and therefore remains
unchanged under every symmetry operation considered here.
The two source masses are interchanged by the generalized reflection,
$M_L \longmapsto M_R$,
and may therefore, if desired, be decomposed into symmetric and
antisymmetric combinations,
$M_\pm = (M_R\pm M_L)/2$,
in complete analogy with the source-center displacements and quadrupole
moments.

The third class consists of the effective coupling strength
$\eta_{\rm th}$ for thermal effects.
Let us assume that the potential $\Phi_{\rm th}$ entering Eq.~\eqref{etaTHdef}
is odd under reflection. This would be the case if,
for example, $\Phi_{\rm th}$ is proportional to an 
undetected temperature gradient across 
the length scale of the apparatus.
The parameter vector therefore naturally separates into
\begin{equation}
\mathbm p = \bigl( \mathbm p_{\rm geom},
\mathbm p_{\rm scalar}, \mathbm p_{\rm coup} \bigr),
\end{equation}
where
$\mathbm p_{\rm geom}$ contains the geometry-dependent effects,
$\mathbm p_{\rm scalar}$ describes physical\ scalars,
and $\mathbm p_{\rm coup}$ contains certain coupling strengths
(for our discussion, only a single one, namely, the thermal one),
\begin{subequations}
\begin{align}
\mathbm p_{\rm geom}
=& \; ( \delta_\pm,
Q_{20}^{(\pm)},
\Gamma_{xx}^{(+)}) \,,
\\[1.3ex]
\mathbm p_{\rm scalar}
=& \; ( G, M_L, M_R) \,,
\\[1.3ex]
\mathbm p_{\rm coup} =& \;
( \eta_{\rm th} ).
\end{align}
\end{subequations}
Only the geometrical parameters possess intrinsic generalized-reflection
parity.  The scalar parameters are invariant (in the case of $G$)
or may easily be decomposed into
reflection-even and reflection-odd combinations, whereas the coupling
coefficients inherit their symmetry entirely from the perturbation
operators to which they are attached.

\subsection{Generalized Least-Squares Formulation} 

The approximate reflection symmetry of the experiment leads to the
first-order selection rules discussed at length
in Sec.~\ref{sec3}. We can give the following 
approximate dependences of the components of the 
vector $\mathbm y$ of observables,
\begin{subequations}
\label{fK}
\begin{align}
\mathbm y_{\rm DC} \simeq & \; 
\mathbm f_{\rm DC}(G,\, \delta_-, \Gamma_{xx}^{(+)}, \, Q^{(+)}_{20}) \,,
\\
\mathbm y_{\rm LC} \simeq & \; 
\mathbm f_{\rm LC}(G,\, \delta_-, \Gamma_{xx}^{(+)}, \, Q^{(+)}_{20}) \,,
\\
\mathbm y_{\rm N} \simeq & \; 
\mathbm f_{\rm N}(\delta_+, \, \eta_{\rm th}) \,,
\\
\mathbm y^{{\rm rot}}_{\rm DC} \simeq & \; 
\mathbm f^{{\rm rot}}_{\rm DC}(G,\, \delta_-, \, \Gamma_{xx}^{(+)}, \, Q^{(+)}_{20}) \,,
\\
\mathbm y^{{\rm rot}}_{\rm LC} \simeq & \; 
\mathbm f^{{\rm rot}}_{\rm LC}(G,\, \delta_-, \, \Gamma_{xx}^{(+)}, \, Q^{(+)}_{20}) \,,
\\
\mathbm y^{{\rm rot}}_{\rm N} \simeq & \;
\mathbm f^{{\rm rot}}_{\rm N}(\delta_+, \, \eta_{\rm th}) \,,
\\
y_{\rm BG} \simeq & \; f_{\rm BG}(\Gamma^{(+)}_{xx}) \,,
\\
y^{\rm rot}_{\rm BG} \simeq & \; f^{{\rm rot}}_{\rm BG}(\Gamma^{(+)}_{xx}) \,,
\end{align}
\end{subequations}
where the details of the $\mathbm f$ functions can only be
determined once the experimental geometry is completely determined.
Residual dependences on the full list of physical 
parameters of our model
are incorporated in the general functional dependence
\begin{equation}
\label{f_with_error}
\mathbm y = \mathbm f(\mathbm p) + \boldsymbol\varepsilon,
\end{equation}
where $\boldsymbol\varepsilon$
denotes the measurement error vector.
The determination of Newton's gravitational constant 
can then be formulated as a
generalized least-squares adjustment.
Suppose that the experiment produces a vector of $m$ observables,
\begin{equation}
\mathbm y = (y_1,y_2,\ldots,y_m)^T,
\end{equation}
as a general form for Eq.~\eqref{full_observables}
while the physical model depends upon a parameter vector
\begin{equation}
\mathbm p = (p_1,p_2,\ldots,p_n)^T,
\end{equation}
as a general form for Eq.~\eqref{full_parameters}.
The covariance matrix of the observations 
(not of the parameters!) is
\begin{equation}
\label{Vdef}
\mathbm V = \mathbb E(\boldsymbol\varepsilon
\otimes \boldsymbol\varepsilon^T) \,,
\end{equation}
where $\mathbb E$ denotes the ensemble average.
Unlike many traditional analyses, the observations are not assumed to
be statistically independent.
Correlations naturally arise because several observables are extracted
from the same interferometric data or because identical calibration
runs contribute simultaneously to several fitted quantities.
The generalized least-squares estimate minimizes
merit function
\begin{equation}
\label{merit1}
\chi^2 = (\mathbm y-\mathbm f)^T V^{-1} (\mathbm y-\mathbm f).
\end{equation}
The use of the inverse covariance matrix as statistical weight ensures
that correlated observations are treated consistently.

The adjustment simultaneously estimates all components of the parameter
vector $\mathbm p$, which are components of 
Eq.~\eqref{full_parameters}.
No distinction is made between the desired quantity and the nuisance
parameters.
Every parameter enters the adjustment on equal mathematical footing.
The covariance matrix of the parameters of the model
follows from the formula (see Appendix~F of Ref.~\cite{MoTa2000}) 
\begin{equation}
\mathbm C = (\mathbm A^T \, \mathbm V^{-1} \, \mathbm A)^{-1} \,,
\end{equation}
where $\mathbm A$ is the Jacobian matrix defined in 
Eq.~\eqref{JACdef} and $\mathbm V$ is the covariance
matrix of the observables defined in Eq.~\eqref{Vdef}.
The uncertainties of the entries $p_i$ of the parameter
vector $\mathbm p$ follow from the diagonal entries
of the covariance matrix $\mathbm C$ as
\begin{equation}
u(p_i) = \sqrt{ (\mathbm C)_{ii} } \,,
\end{equation}
where no summation over $i$ is carried out.
The first entry of the vector $\mathbm p$ is the 
gravitational constant $G$ itself; hence, once the 
least-squares analysis has been carried out,
the best estimate for $G$ is obtained from the 
entry in the first row, and first column, of the 
covariance matrix of the parameters.

%
% Generalized Least--Squares Adjustment
%
\section{Generalized Least--Squares Adjustment}
\label{sec5}

%
% Model
%
\subsection{Functional Dependence of the Model}
\label{sec5A}

We shall consider the application of a generalized
least-squares adjustment to the modeling 
of a vector of observables $\mathbm y$ by
a given set of parameters $\mathbm p$.
For reference, it is useful to recall, 
from Eq.~\eqref{full_observables}, the full vector
of observables that we had identified 
for the space-based determination of the gravitational constant, 
\begin{equation}
\mathbm y = \left( \mathbm y_{\rm DC}, \mathbm y_{\rm LC}, \mathbm y_{\rm N},
\mathbm y^{{\rm rot}}_{\rm DC}, \mathbm y^{{\rm rot}}_{\rm LC},
\mathbm y^{{\rm rot}}_{\rm N}, y_{\rm BG}, y^{{\rm rot}}_{\rm BG} \right)^T .
\end{equation}
with the entries representing the 
direct channel (DC), the lock-in channel (LC),
the null coordinate (N), the rotated versions thereof (rot),
and background coordinates recorded with the 
source masses absent or very far away.
We also recall the vector of parameters,
from Eq.~\eqref{full_parameters},
\begin{equation}
\mathbm p = \left( G, M_L, M_R, \delta_+, \delta_-, 
Q_{20}^{(+)}, Q_{20}^{(-)}, 
\Gamma_{xx}^{(+)}, \eta_{\rm th} \right)^T,
\end{equation}
with the gravitational constant $G$, the source
masses $M_L$ and $M_R$, the symmetry-dependent
source-mass displacements $\delta_\pm$, 
the symmetric and antisymmetric components of the 
quadrupole term $Q_{20}^{(\pm)}$,
the gradient $\Gamma_{xx}^{(+)}$, 
and the thermal coefficient $\eta_{\rm th}$.
The general functional dependence is 
given in Eq.~\eqref{f_with_error},
$\mathbm y = \mathbm f(\mathbm p) + \boldsymbol\varepsilon$,
where $\boldsymbol\varepsilon$
denotes the measurement error vector.
A central assumption is that, if we take the 
correct parameter $\mathbm p_{\rm true}$ and 
plug them into the function $\mathbm f$,
then we obtain the ensemble average $\mathbb E$
of the observables,
\begin{equation}
\mathbb E(\mathbm y) = 
\mathbm f(\mathbm p_{\rm true}) + 
\mathbb E( \boldsymbol\varepsilon ) \,,
\quad
\mathbb E( \boldsymbol\varepsilon )  = \mathbm 0 \,.
\end{equation}
In the following, we shall discuss
some general aspects of the generalized 
least-squares problem which consists
in finding an optimal approximation 
$\widehat{\mathbm p} \simeq \mathbm p_{\rm true}$
to the true parameters designed so that
$\widehat{\mathbm p}$ approximates, in an 
optimal way, the weighted mean $\mathbm y$ 
of the observables obtained from data taking
within a given experiment (or, experiments).

We shall consider a general model with $m$ observables
and $n$ parameters, overdetermined, so that $m > n$.
In order to fix ideas, we mention certain conventions
that are used in the following.
Nature is assumed to produce individual realizations $\mathbm y^{(k)}$
in the $k$th iteration of the measurements.
The experiment averages them to obtain the best estimate $\overline{\mathbm y}$.
The generalized least-squares adjustment infers the underlying physical parameters
$\widehat{\mathbm p}$.
The observation model maps those fitted parameters back into the observable space as
$\widehat{\mathbm y} = {\mathbm f}(\widehat{\mathbm p})$.
The optimum estimation of the observables, obtained
from the optimum parameters (prediction of the fitted model),
is $\widehat{\mathbm y} = {\mathbm f}(\widehat{\mathbm p})$.
At the optimum, we observe that the residual vector
(difference of averaged observations and fitted model)
is $\mathbm r = \overline {\mathbm y} - \widehat{\mathbm y}$.

%
% Nonlinear Observation Model
%
\subsection{Nonlinear Observation Model}
\label{sec5B}

The optimization procedure is different for 
nonlinear compared to linear models.
We first discuss the more general nonlinear case.
The superscript $n$ used in this section labels successive
\emph{iterations of the parameter adjustment}.  It should not be
confused with the index $k$, which labels repeated experimental
data-taking events.
The experimentally combined observation vector is denoted by
$\overline{\mathbm y}$ and its covariance matrix by
$\bm V = \bm V_{\overline{\bm y}}$. 

For a valid observation model,
one starts with the assumption that the model
encoded in the $\bm f$ function is able to 
describe the observations, up to some
observational error $\overline{\boldsymbol{\varepsilon}}_{\rm obs}$
which affects experimental results for a 
finite set of taken data. The presence of the 
experimental error $\overline{\boldsymbol{\varepsilon}}_{\rm obs}$.
implies that the best estimate
$\overline{\mathbm y}$ for the observables,
obtained from data taking (with a necessarily
number of observations), deviates from the 
model $\bm f(\bm p_{\rm true})$	evaluated at the 
true parameter values $\bm p_{\rm true}$.
We set
\begin{equation}
\overline{\mathbm y} =
\bm f(\bm p_{\rm true})
+ \overline{\boldsymbol{\varepsilon}}_{\rm obs},
\end{equation}
where \(\overline{\boldsymbol{\varepsilon}}_{\rm obs}\) is the particular
realized mean observational error of the finite data set.  Although,
when averaged over the entire statistical ensemble,
$\mathbb E[ \overline{\boldsymbol{\varepsilon}} ] = \mathbm 0$,
the realized vector
\(\overline{\boldsymbol{\varepsilon}}_{\rm obs}\) is generally not
identically zero.

The nonlinear adjustment begins with an initial estimate
$\mathbm p^{(n=0)}$.
The observation model evaluated at this parameter vector is
$\mathbm f(\mathbm p^{(0)})$,
and the corresponding residual vector is
$\mathbm r^{(0)} = \overline{\mathbm y} - \mathbm f(\mathbm p^{(0)})$.
More generally, at iteration \(n\),
one uses a parameter estimate $\mathbm p^{(n)}$ with
and residuals $\mathbm r^{(n)}$ with
\begin{equation}
\mathbm r^{(n)} = 
\overline{\mathbm y} - \mathbm f(\mathbm p^{(n)}) = 
\bm f(\bm p_{\rm true}) - \mathbm f(\mathbm p^{(n)}) +
\overline{\boldsymbol{\varepsilon}}_{\rm obs} \,.
\end{equation}
In the $n$th step,
the nonlinear observation model is expanded to first order about the
current parameter estimate,
\begin{equation}
\mathbm f( \mathbm p^{(n)} + \Delta\mathbm p )
\simeq
\mathbm f(\mathbm p^{(n)}) + A^{(n)}\Delta\mathbm p,
\end{equation}
where
\begin{equation}
( A^{(n)} )_{ij} = \left.  \frac{\partial f_i}{\partial p_j}
\right|_{\mathbm p=\mathbm p^{(n)}}
\end{equation}
is the Jacobian matrix evaluated at the current parameter estimate
($i = 1,\dots,m$ and $j=1,\dots,n$).
The locally linearized residual is therefore
\begin{equation}
\overline{\mathbm y} - \mathbm f( \mathbm p^{(n)}+\Delta\mathbm p )
\simeq
\mathbm r^{(n)} - \mathbm A^{(n)} \, \Delta\mathbm p \,.
\end{equation}
In the $n$th iteration,
the parameter correction $\Delta\mathbm p$
is chosen to minimize
\begin{equation}
\chi_n^2
= [ \mathbm r^{(n)} - \bm A^{(n)} \, \Delta\mathbm p ]^T 
{\bm V}^{-1} 
[ \mathbm r^{(n)} - \bm A^{(n)} \, \Delta\mathbm p ].
\end{equation}
After differentiation with respect to \(\Delta\mathbm p\), 
and setting the result equal to zero,
one obtains the local normal equations,
\begin{equation}
(\bm A^{(n)})^T \, {\bm V}^{-1} \, {\bm A}^{(n)}
\, \Delta\widehat{\mathbm p}^{(n)}
= (\bm A^{(n)})^T \, \bm V^{-1} \mathbm r^{(n)}.
\end{equation}
Provided that the local normal matrix is nonsingular,
one obtains the iterative shift $\Delta\widehat{\mathbm p}^{(n)}$
of the parameters as follows,
\begin{equation}
\label{iterative_shift}
\Delta\widehat{\mathbm p}^{(n)}
= [ (\bm A^{(n)})^T \, \bm V^{-1} \,
\bm A^{(n)} ]^{-1} \, (\bm A^{(n)})^T \, \bm V^{-1} \,
\mathbm r^{(n)}.
\end{equation}
The parameter vector is then updated according to
\begin{equation}
\mathbm p^{(n+1)} = \mathbm p^{(n)} +
\Delta\widehat{\mathbm p}^{(n)}.
\end{equation}
The observation model and Jacobian are subsequently recomputed at the
new parameter estimate, 
\begin{subequations}
\begin{align}
\bm r^{(n+1)} =& \;
 \overline{\bm y} - \mathbm f(\mathbm p^{(n+1)}),
\\
( A^{(n+1)} )_{ij} =& \;
\left.  \frac{\partial f_i}{\partial p_j}
\right|_{\mathbm p=\mathbm p^{(n+1)}} \,.
\end{align}
\end{subequations}
The sequence of operations is therefore
$\mathbm p^{(n)} \longmapsto
\mathbm f(\mathbm p^{(n)}) \longmapsto
\bm A^{(n)} \longmapsto
\Delta\widehat{\mathbm p}^{(n)}
\longmapsto \mathbm p^{(n+1)}$.
We assume that, in the limit $n \to \infty$,
this process converges to 
\begin{align}
\widehat{\bm p} =& \; \bm p^{(\infty)} 
\equiv \lim_{n \to \infty} \bm p^{(n)} \,,
\\
\widehat{\bm A} =& \; \bm A^{(\infty)} 
\equiv \lim_{n \to \infty} \bm A^{(n)} \,,
\end{align}
This is the Gauss--Newton form of the nonlinear generalized
least-squares adjustment and is closely analogous to the classical
Newton--Raphson iteration.
The iteration is continued until the parameter corrections become
negligible.  One possible componentwise criterion is
\begin{equation}
\left| \Delta\widehat p_i^{(n)} \right|
\ll u(\widehat p_i)
\end{equation}
for every adjusted parameter $p_i$.
Here, $u(\widehat p_i)$ is the uncertainty
assigned to the parameter $p_i$, to be determined
using the method of covariances, with 
details as described in the following.
Equivalently, one may require that successive changes in
\(\chi^2\), in the parameter vector, and in the predicted observations
all fall below prescribed numerical tolerances.
At the point of convergence, for $n \to \infty$, we have
\begin{equation}
{\bm A}^{(\infty)} \, {\bm V}^{-1} \,
\left[ \overline{\mathbm y} - \mathbm f(\widehat{\mathbm p}) \right]
= \mathbm 0,
\end{equation}
where the Jacobian is evaluated at the 
point of convergence,
\begin{equation}
( A^{(\infty)} )_{ij} = \left.  \frac{\partial f_i}{\partial p_j}
\right|_{\mathbm p=\mathbm p^{(\infty)}} \,.
\end{equation}
Thus, the final residual 
(after convergence is reached) is orthogonal, in the 
$\mathbm V^{-1}$-weighted
scalar product, to the tangent space of the nonlinear model manifold at
the fitted point. One notes that, at the 
point of convergence, 
the fitted model prediction for the observables is
\begin{equation}
\widehat{\bm y} = \bm f( \widehat{\bm p} ) \,.
\end{equation}
The final residuals 
\begin{equation}
\bm r^{(\infty)} = \widehat{\bm r} = 
\overline{\mathbm y} - \mathbm f(\widehat{\mathbm p})
\end{equation}
are not necessarily vanishing.

%
% Linear Model
%
\subsection{Linear Observation Model}
\label{sec5C}

Now consider the exactly linear observation model
where $f(\bm p) = \mathbm A \, \mathbm p$.
This is a good approximation in cases where 
the entries of $\bm p$ are known to numerically small
quantities. We set
\begin{equation}
\overline{\mathbm y} = \bm A \, \mathbm p_{\rm true}
+ \overline{\boldsymbol{\varepsilon}}_{\rm obs},
\end{equation}
where $\mathbm A$ is a constant matrix 
and does not depend on the parameter vector.
One begins with an arbitrary initial estimate which 
can be chosen as
\begin{equation}
\mathbm p^{(0)} = \bm 0.
\end{equation}
The initial residual is
\begin{equation}
\mathbm r^{(0)}
= \overline{\mathbm y} - \bm A \, \mathbm p^{(0)}
= \overline{\mathbm y} \,.
\end{equation}
The first generalized least-squares correction is
\begin{equation}
\Delta\widehat{\mathbm p}^{(0)}
= \left( \bm A^T \, \bm V^{-1} \, \bm A \right)^{-1} 
\bm A^T \, \bm V^{-1} \, \overline{\mathbm y} \,.
\end{equation}
Consequently,
\begin{equation}
\label{p1}
\mathbm p^{(1)} = \widehat{\mathbm p} =
\left( \bm A^T \, \bm V^{-1} \, \bm A \right)^{-1}
\bm A^T \, \bm V^{-1} \, \overline{\mathbm y}.
\end{equation}

We will now convince ourselves that the 
second correction to the parameters vanishes.
After the first update, the residual is
\begin{equation}
\mathbm r^{(1)} = \overline{\mathbm y} - \bm A 
\widehat{\mathbm p}.
\end{equation}
According to Eq.~\eqref{iterative_shift},
the next correction is therefore
\begin{multline}
\Delta\widehat{\mathbm p}^{(1)}
= \left( \bm A^T \, \bm V^{-1} \, \bm A \right)^{-1} \,
\bm A^T \, \bm V^{-1} \, \mathbm r^{(1)}
\\
= \left[ \left( \bm A^T \, \bm V^{-1} \, \bm A \right)^{-1} \, 
\bm A^T \, \bm V^{-1} \, \bm y \right] 
- \left( \bm A^T \, \bm V^{-1} \, \bm A \right)^{-1} \, 
( \bm A^T \, \bm V^{-1} \, \bm A ) \, 
\widehat{\mathbm p} = \mathbm 0.
\end{multline}
For the first term, we use Eq.~\eqref{p1},
while for the second term, we use the fact that 
\begin{equation}
\left( \bm A^T \, \bm V^{-1} \, \bm A \right)^{-1} \, 
( \bm A^T \, \bm V^{-1} \, \bm A ) = \bm 1 \,.
\end{equation}
Alternatively, we observe that the 
residuals $\mathbm r^{(1)}$ fulfill the 
normal equations at the generalized least-squares optimum,
\begin{equation}
\bm A^T \, \bm V^{-1} \, \mathbm r^{(1)}
= \mathbm 0 \,,
\qquad
\mathbm r^{(\infty)} = \widehat{\bm r} = \mathbm r^{(1)} \,.
\end{equation}
The fact that the parameter correction vanishes after one step does not
imply that the fitted residual itself vanishes.
In the linear case, the formula for the 
experimental observables obtained by the fitted
model assumes a particularly simple form,
\begin{equation}
\widehat{\bm y} = \bm A \, \widehat{\bm p} \,.
\end{equation}
In an overdetermined experiment with observational noise,
the residuals are
\begin{equation}
\widehat{\mathbm r} = \overline{\mathbm y} - \bm A\widehat{\mathbm p}
\end{equation}
Thus no further change of the adjusted parameters can reduce the
weighted residual norm to first order.

%
% Covariances of the Parameters
%
\subsection{Covariances of the Parameters}
\label{sec5D}

Of course, our final goal is to assign uncertainties
to the parameters $\bm p$.  The parameters $\bm p$
take the same role as the fundamental constants 
of physics in the CODATA least-squares adjustment~\cite{MoTa2000}.
They constitute the entries in the parameter vector $\bm p$,
the first of which is the gravitational constant.
The covariance matrix $\widehat{\bm C}$ of the 
estimated parameters is obtained 
from the covariance matrix $\bm V$ of the observations 
via the Jacobian, at the point of convergence 
\begin{equation}
\widehat{\mathbm C} = (
\widehat{\mathbm A}^T \, 
\mathbm V^{-1} \, 
\widehat{\mathbm A})^{-1},
\end{equation}
where $\widehat{\mathbm A}$, in the case of a
nonlinear model, is the Jacobian evaluated in the 
final (converged) iteration of the parameter update.
For the linear model, since $\bm A$ is constant, we 
can write
$\mathbm C = (
\mathbm A^T \,
\mathbm V^{-1} \,
\mathbm A)^{-1}$.
The diagonal elements $(\mathbm C)_{ii}$ of $\mathbm C$
represent the variances of the adjusted parameters,
\begin{equation}
u^2(p_i) = (\widehat{\mathbm C})_{ii},
\end{equation}
while the off-diagonal elements describe their statistical
correlations (the Einstein convention is not used).
The corresponding correlation coefficients 
(of the parameters) are given as
\begin{equation}
\rho_{ij} = \frac{(\widehat{\mathbm C})_{ij}}
{\sqrt{(\widehat{\mathbm C})_{ii} \, 
(\widehat{\mathbm C})_{jj}}} \,,
\end{equation}
where, again, the Einstein summation convention is not 
applied. The numerical values of these covariances depend upon the statistical
properties of the experimental observations and therefore cannot be
evaluated universally.

We briefly summarize
the distinction between the linear and nonlinear cases.
In the linear case, we have
$\mathbm f(\mathbm p)= \bm A \, \mathbm p$
with constant $\bm A$;
one generalized least-squares update is sufficient.
For the nonlinear case, with 
$\mathbm y = \mathbm f(\mathbm p)$,
repeated Gauss--Newton updates are required.

%
% Illustrative Example
%
\section{Illustrative Example}
\label{sec6}

%
% Illustrative Observation Vector
%
\subsection{Illustrative Observation Vector}
\label{sec6A}

In order to illustrate the application of the procedure for the least-squares
adjustment, we shall consider an abridged version of the model outlined above
in Eqs.~\eqref{full_observables},~\eqref{full_parameters}
and~\eqref{f_with_error}.  For clarity, we shall choose an illustrative
covariance model which is linear; in this case, as outlined in
Sec.~\ref{sec5B}, the optimization of the parameters can be carried out in one
step.  Our illustrative model will be simplified, yet physically motivated.
Although the numerical values introduced in the present section are purely
illustrative, the structure of the model reflects the principal features of the
proposed experiment.  The purpose of the model is not to predict the final
mission accuracy, but rather to demonstrate how multiple science channels, null
measurements, and dedicated calibration measurements can be combined within a
unified covariance analysis.
The dimensionless observation vector is chosen as
\begin{equation}
\label{model_observables}
\mathbm y = \frac{1}{y_{\rm ref}} \,
(y_{\rm LC}^{[1]}, y_{\rm LC}^{[2]},
y_{\rm LC}^{[1]}, y_{\rm LC}^{[2]},
y_{\rm N}, y_{\rm LC}^{\rm rot}, y_{\rm BG} )^T \,,
\end{equation}
where $y_{\rm ref}$ is a suitable reference scale
for typical accelerations recorded in the apparatus,
and the six observables have the following interpretation.
[For source masses of the order of one tonne, and
apparatus dimensions of the order of one meter,
a possible choice,
with reference to Eq.~\eqref{AG},
is $y_{\rm ref} = 10^{-10} \, {\rm m}/{\rm s}^2$.]
First, $y_{\rm DC}$ constitutes the 
acceleration (proportional to $G$) in the direct
channel. The variables $y_{\rm LC}^{(1)}$ and
$y_{\rm LC}^{(2)}$ denote lock-in observables obtained for two
different source-mass separations. Since the corresponding geometrical
configurations differ, the two observables possess different
sensitivities to the adjusted parameters and therefore provide
additional leverage in the least-squares adjustment.
The observable $y_{\rm N}$ denotes the reflection-sensitive null observable.
Ideally, this quantity vanishes by symmetry and therefore provides a
sensitive diagnostic for reflection-odd perturbations.
The variable $y_{\rm LC}^{\rm rot}$ denotes a lock-in
measurement performed after rotating the experimental apparatus by
$90^\circ$, and after subtracting the 
gravitationally induced motion 
for the unrotated source masses. This observable is designed primarily to constrain the
reflection-even quadrupole parameter.
The background observable $y_{\rm BG}$ denotes the background null observable 
obtained after removing the source masses (or moving them far away). 
It therefore provides an independent
estimate of the instrumental background and possible residual offsets.

\subsection{Illustrative Parameter Vector}
\label{sec6B}

The illustrative adjusted parameter vector is chosen as
\begin{equation}
\label{model_parameters}
\mathbm p = (\mathcal G, \delta_+, \delta_-, q, \gamma)^T,
\end{equation}
where all parameters are dimensionless.
and numerically small (which facilitates the 
application of the linear model).
The dimensionless entries in the parameter vector~\eqref{model_parameters}
therefore differ from those in Eq.~\eqref{full_parameters}.
The first entry, 
\begin{equation}
\label{calGdef}
\mathcal G = \frac{\Delta G/G_0}{G_{\rm ref}},
\end{equation}
denotes the relative deviation $\Delta G = G - G_0$ 
of the gravitational constant from a reference value, where the 
reference value $G_0$ could be chosen numerically
close to the current CODATA value for $G$
(see Ref.~\cite{MohrEtAl2025}), and $G_{\rm ref}$ is a reference scale.
The parameters $\delta_+$ and $\delta_-$ 
describe the relative displacement of the source
masses from their nominal reference positions,
\begin{equation}
\delta_+ = \frac{d_+}{d_{\rm ref}},
\qquad
\delta_- = \frac{d_-}{d_{\rm ref}},
\end{equation}
where $d_{\rm ref}$ is a reference scale, and
we refer to Fig.~\ref{fig1} for the definition of $d_\pm$.
The dimensionless parameter
\begin{equation}
q = \frac{Q_{20}^{(+)}}{Q_{\rm ref}},
\end{equation}
parameterizes the reflection-symmetric component of the 
quadrupoles of the source masses ($Q_{\rm ref}$ is a reference scale).
The symmetric component of the gradient of the tidal gravitational 
field is described by
\begin{equation}
\gamma = \frac{\Gamma_{xx}^{(+)}}{\Gamma_{\rm ref}},
\end{equation}
where $\Gamma_{\rm ref}$ is a reference scale.

Here \(G_0\) denotes the adopted reference value of the gravitational
constant, while \(d_{\rm ref}\), \(Q_{\rm ref}\), and
\(\Gamma_{\rm ref}\) are convenient normalization constants. Their
precise numerical values are irrelevant for the illustrative model and
merely serve to place all adjusted parameters on comparable numerical
scales.

Reflecting a relatively modest demand on the accuracy of the 
experiment, one could assume the following values 
for the reference scales (again, purely for illustrative 
purposes),
\begin{subequations}
\begin{align}
\label{refGdef}
G_{\rm ref} =& \; 10^{-6},
\\
d_{\rm ref} =& \; 1~\mu{\rm m},
\\
Q_{\rm ref} =& \; 10^{-2}~{\rm kg\,m^2},
\\
\Gamma_{\rm ref} =& \; 10^{-12}~{\rm s^{-2}} \, .
\end{align}
\end{subequations}
These values are order-of-magnitude normalization choices rather than
elements of the mission error budget. For source 
masses with a radial dimension of the order of 
$R \sim 1\,$m, the displacement scale follows
from $d_{\rm ref}/R\sim10^{-6}$, while the quadrupole scale follows
approximately from $Q_{\rm ref}/(MR^2) \sim 10^{-6}$.
The gravity-gradient scale corresponds to a differential acceleration
of order $10^{-12}~{\rm m\,s^{-2}}$ across a baseline of approximately
one meter.

%
% Statistical Picture
%
\subsection{Statistical Picture}
\label{sec6C}

In order to fix ideas, we shall, once more,
dwell the linear statistical model discussed 
in Sec.~\ref{sec5C}. Let us remember 
that, in practice, the observation vector entering the generalized
least-squares adjustment is generally not obtained from a single
experimental realization. Rather, each observable is determined from a
large number of repeated measurements.
Let
\begin{equation}
\mathbm y^{(k)}, \qquad k=1,\ldots,N,
\end{equation}
denote the observation vector obtained during the $k$th
experimental realization.

Each realization satisfies the linear observation model
[see also Eq.~\eqref{f_with_error}]
\begin{equation}
\mathbm y^{(k)} = \mathbm A \, \mathbm p_{\rm true} +
\boldsymbol{\varepsilon}^{(k)},
\end{equation}
where
\begin{equation}
\mathbm p_{\rm true} = (\mathcal G_{\rm true}, 
d_{+,{\rm true}}, d_{-,{\rm true}}, 
q_{\rm true}, \gamma_{\rm true})^T
\end{equation}
is the true parameter vector and
$\boldsymbol{\varepsilon}^{(k)}$
is the corresponding observation-error vector.
The observation errors are assumed to satisfy
$\mathbb E(\boldsymbol{\varepsilon}^{(k)}) = \mathbm 0$,
and the covariance matrix for a single 
measurement is
\begin{equation}
{\rm cov}( \varepsilon^{(k)}_i, \varepsilon^{(k)}_j )
= (\mathbm V_{\rm single})_{ij} \,,
\end{equation}
when expressed in terms of its 
matrix elements ($i,j = 1,\dots,5$).

The experimentally determined observation vector entering the
generalized least-squares adjustment is the sample mean
\begin{equation}
\label{sample_mean}
\overline{\mathbm y} = \frac{1}{N}
\sum_{k=1}^{N} \mathbm y^{(k)}.
\end{equation}
Its expectation value is
$\mathbb E(\overline{\mathbm y}) = 
\mathbm A \, \mathbm p_{\rm true}$.
Assuming statistically independent repetitions,
the covariance matrix of the averaged observations is
\begin{equation}
\label{Vsingle}
\mathbm V_{\overline{\mathbm y}} = \mathbm V_{\rm single}/N \,.
\end{equation}
Within the linear model,
the generalized least-squares estimate of the parameter vector is
\begin{equation}
\widehat{\mathbm p} = \left( \mathbm A^T \, 
\mathbm V_{\overline{\mathbm y}}^{-1} \, \mathbm A \right)^{-1}
\mathbm A^T \, \mathbm V_{\overline{\mathbm y}}^{-1}
\overline{\mathbm y}.
\end{equation}
Substituting Eqs.~\eqref{sample_mean} and \eqref{Vsingle}, one finds
that the common factor \(N\) cancels,
\begin{equation}
\widehat{\mathbm p} = 
\left( {\mathbm A}^T \, {\mathbm V}_{\rm single}^{-1} \,
{\mathbm A} \right)^{-1} {\mathbm A}^T \, {\mathbm V}_{\rm single}^{-1} \,
\overline{\mathbm y}.
\end{equation}
Thus, averaging the observations changes neither the form of the
generalized least-squares estimator nor its central value.
Taking the expectation value for entire 
statistical ensemble, one finds that 
\begin{equation}
\mathbb E( \widehat{\mathbm p} )
=
\left( {\mathbm A}^T \, 
{\mathbm V}_{\mathbb E(\overline{\mathbm y})}^{-1} \,
{\mathbm A} \right)^{-1} {\mathbm A}^T \, 
{\mathbm V}_{\mathbb E(\overline{\mathbm y})}^{-1} \,
\mathbb E( \overline{\mathbm y} )
= \left( {\mathbm A}^T \, 
{\mathbm V}_{\mathbb E(\overline{\mathbm y})}^{-1} \,
{\mathbm A} \right)^{-1} ( {\mathbm A}^T \, 
{\mathbm V}_{\mathbb E(\overline{\mathbm y})}^{-1} \,
{\mathbm A} ) \, \mathbm p_{\rm true}
= \mathbm p_{\rm true}.
\end{equation}
For an ideal experiment,
\begin{equation}
\mathbm p_{\rm true}
= ({\mathcal G}_{\rm true},0,0,0,0)^T \,.
\end{equation}
Using 
\begin{equation}
\mathbb E(\widehat{\mathbm p})
= (\mathbb E(\widehat {\mathcal G}),0,0,0,0)^T \,,
\end{equation}
one finds that, provide the model
describes the experiment accurately, the ensemble mean of the
best estimate for $\mathcal G$ equals its true value,
$\mathbb E(\widehat{\mathbm p}) = \widehat{\mathcal G}_{\rm true}$.

The fitted parameter vector predicts the observation vector
$\widehat{\mathbm y} = {\mathbm A} \, \widehat{\mathbm p}$,
which should not be confused with the experimentally averaged
observation vector $\overline{\mathbm y}$.
The residual vector is therefore
$\mathbm r = \overline{\mathbm y} - \widehat{\mathbm y} =
\overline{\mathbm y} - {\mathbm A} \, \widehat{\mathbm p}$.

\subsection{Illustrative Vectors and Matrices}
\label{sec6D}

Although the numerical values of the Jacobian depend upon the detailed
geometry of the experiment, its qualitative structure follows directly
from the symmetry considerations developed in Sec.~\ref{sec3}.
For the six illustrative observables given in
Eq.~\eqref{model_observables} and the five illustrative
parameters given in Eq.~\eqref{model_parameters},
the corresponding illustrative $7 \times 5$ Jacobian 
matrix is chosen as
\begin{equation}
\mathbm A=
\begin{pmatrix}
1.00 & 0 & 2.00 & 0.10 & \kappa^{[1]}_{\rm DC}
\\[1ex]
0.50 & 0 & 0.80 & 0.10 & \kappa^{[2]}_{\rm DC}
\\[1ex]
1.00 & 0 & 4.00 & 0.20 & \kappa_{\rm LC}^{[1]}
\\[1ex]
0.40 & 0 & 0.40 & 0.10 & \kappa_{\rm LC}^{[2]}
\\[1ex]
0 & 1.00 & 0 & 0 & \varepsilon_{\rm N}
\\[1ex]
0 & 0 & 0.05 & 1.00 & 0
\\[1ex]
0 & 0 & 0 & 0 & \varepsilon_{\rm BG}
\end{pmatrix}.
\end{equation}
The rows correspond to the observables,
the columns correspond to the parameters.
The coefficients $\kappa_{\rm DC}$,
$\kappa_{\rm LC}^{(1)}$ and $\kappa_{\rm LC}^{(2)}$
represent symmetry-allowed first-order sensitivities of the
corresponding observables to the reflection-even gravity-gradient
parameter; they are still expected to be numerically
suppressed in view of the smallness of the tidal 
gravity gradient.
By contrast, $\varepsilon_{\rm N}$ and $\varepsilon_{\rm BG}$
parameterize residual symmetry-breaking leakages into the null and
background channels. In the ideal reflection-symmetric apparatus these
coefficients vanish identically.
For the illustrative model, we adopt
the following numerical values of the $\kappa$ and 
$\epsilon$ parameters,
\begin{subequations}
\begin{align}
\kappa_{\rm DC}^{[1]} =& \;  \kappa_{\rm DC}^{[2]} = 0.02, \\
\kappa_{\rm LC}^{[1]} =& \;  0.017 \,, \\
\kappa_{\rm LC}^{[2]} =& \;  0.013 \,, \\
\varepsilon_{\rm N} =& \; \varepsilon_{\rm BG} = 0.007.
\end{align}
\end{subequations}
The equality of the small coefficients is introduced solely for
illustrative simplicity. In a realistic mission analysis these
quantities would be determined by detailed instrument modelling and
would generally assume different numerical values.

We take the following averaged result for the 
observables from the experiment 
\begin{equation}
\overline{\boldsymbol{y}} = 
(0.12,\, 0.07, 0.14,\, 0.09,\, 0.01,\, 0.20,\, 0.01)^T.  
\end{equation}
For our model calculation,
the illustrative standard uncertainties (associated 
with the experimental data) are assumed to be 
\begin{equation}
\boldsymbol{\sigma}_{\overline{\boldsymbol{y}}} = 
(0.01,\, 0.01, \, 0.03, \, 0.03,\, 0.10,\, 0.02,\, 0.01)^T.  
\end{equation}
In the statistical sense, referencing 
data taking, the entries of the 
$\mathbm \sigma$ vector are to be obtained 
from the standard deviations of the 
best estimates of the observables.
So, the first entry of the $\mathbm \sigma$ 
vector corresponds to the uncertainty 
$u(\overline{\mathbm y}_{\rm DC})/y_{\rm ref}$ 
of the mean value of the DC channel acceleration observable
(divided by the reference scale $y_{\rm ref}$).
We assume the following illustrative
values for the correlation matrix $\mathbm R$
of the observations,
\begin{equation}
\mathbm R_{\overline{\mathbm y}} =
\begin{pmatrix}
1 & 0.30 & 0.20 & 0.10 & 0 & 0 & 0
\\[1ex]
0.30 & 1 & 0.20 & 0.10 & 0 & 0 & 0
\\[1ex]
0.20 & 0.20 & 1 & 0.10 & 0 & 0 & 0
\\[1ex]
0.10 & 0.10 & 0.10 & 1 & 0 & 0 & 0 
\\[1ex]
0 & 0 & 0 & 0 & 1 & 0 & 0.30
\\[1ex]
0 & 0 & 0 & 0 & 0 & 1 & 0
\\[1ex]
0 & 0 & 0 & 0 & 0.30 & 0 & 1
\end{pmatrix}.
\end{equation}
The observation covariance matrix is obtained from
\begin{equation}
(\mathbm V_{\overline{\mathbm y}})_{ij} = 
(\mathbm \sigma_{\overline{\boldsymbol{y}}} )_i \, 
(\mathbm \sigma_{\overline{\boldsymbol{y}}} )_j \, 
(\mathbm R_{\overline{\mathbm y}})_{ij} \,.
\end{equation}
Again, 
the numerical values adopted above are intended only to illustrate the
operation of the generalized least-squares adjustment. They should not
be interpreted as representing the expected uncertainty budget of the
proposed experiment.

For reference, we recall that the 
covariance matrix is obtained as
\begin{equation}
\mathbm C_{\overline{\mathbm y}} = ( \mathbm A^T \, 
\mathbm V_{\overline{\mathbm y}} \, \mathbm A )^{-1} \, .
\end{equation}
The correlation matrix $\mathbm \rho$
of the parameters (not observations!), whose elements are
\begin{equation}
( \mathbm \rho_{\overline{\boldsymbol{y}} } )_{ij}
= \frac{ ( \mathbm C_{\overline{\mathbm y}} )_{ij} }%
{ \sqrt{ ( \mathbm C_{\overline{\mathbm y}} )_{ii} \;
( \mathbm C_{\overline{\mathbm y}} )_{jj}  } } \,,
\end{equation}
is given as
\begin{equation}
\mathbm \rho_{\overline{\boldsymbol{y}}} =  
\left( \begin{array}{rrrrr}
 1.000  & -0.095 & -0.883 & -0.064 & -0.748 \\
 -0.095 &  1.000 &  0.056 & -0.002 &  0.127 \\
 -0.883 &  0.056 &  1.000 & -0.033 &  0.441 \\
 -0.064 & -0.002 & -0.033 &  1.000 & -0.016 \\
 -0.748 &  0.127 &  0.441 & -0.016 &  1.000 \\
\end{array} \right).
\end{equation}
The fitted model is
\begin{equation}
\label{Lconnect}
\widehat{\mathbm p} = {\mathbm C}_{\overline{\mathbm y}} \,
\mathbm A^T \, \mathbm V_{\overline{\mathbm y}}^{-1} \,
\overline{\mathbm y}  
= (0.0370, -0.0019, 0.0087, 0.1998, 1.2913)^T \,,
\end{equation}
with the following uncertainties,
\begin{equation}
{\mathbm u}(\widehat{\mathbm p}) =
\sqrt{ {\rm diag}( \mathbm C_{\overline{\mathbm y}} ) }
= (0.0427, 0.0962, 0.0145, 0.0200, 0.8713)^T \,,
\end{equation}
where ${\rm diag}( \mathbm C_{\overline{\mathbm y}} )$
is the vector of diagonal elements of 
$\mathbm C_{\overline{\mathbm y}}$ and the 
square root is understood to be applied to all diagonal elements.
From the first entry of the 
$\widehat{\mathbm p}$ vector, one would conclude that
the hypothetical experiment leads to a determination 
of the gravitational constant by 
\begin{equation}
\label{Gresult}
\mathcal G = \frac{\Delta G/G_0}{G_{\rm ref}} =
0.0370 \pm 0.0427 \,,
\end{equation}
where we recall Eqs.~\eqref{calGdef} and~\eqref{refGdef}.

%
% Restricted DC and LC Models
%
\subsection{Restricted DC and LC Models}
\label{sec6E}

One of the main motivations for introducing
the DC and LC channels for the determination of $G$
is the redundancy and the possibility of 
cross-checks between the DC and LC channels. 
In particular, we observe that the 
first and second entries in the vector $\overline{\mathbm y}$
pertain to the DC channel, 
while the third and fourth entries in the
vector $\overline{\mathbm y}$ pertain to the LC channel.
We can thus define the matrices
\begin{equation}
{\mathbm S}_{\rm DC} =
\left( \begin{array}{ccccccc}
 1 & 0 & 0 & 0 & 0 & 0 & 0 \\
 0 & 1 & 0 & 0 & 0 & 0 & 0 \\
 0 & 0 & 0 & 0 & 1 & 0 & 0 \\
 0 & 0 & 0 & 0 & 0 & 1 & 0 \\
 0 & 0 & 0 & 0 & 0 & 0 & 1 \\
\end{array} \right) \,,
\end{equation}
and 
\begin{equation}
{\mathbm S}_{\rm LC} =
\left( \begin{array}{ccccccc}
 1 & 0 & 0 & 0 & 0 & 0 & 0 \\
 0 & 1 & 0 & 0 & 0 & 0 & 0 \\
 0 & 0 & 0 & 0 & 1 & 0 & 0 \\
 0 & 0 & 0 & 0 & 0 & 1 & 0 \\
 0 & 0 & 0 & 0 & 0 & 0 & 1 \\
\end{array} \right) \,,
\end{equation}
which, when applied to the observation vector $\overline{\mathbm y}$,
select the variables relevant to the DC and LC channels
(we keep the rotated sixth entry of $\overline{\mathbm y}$
for reference purposes in both separate DC and LC calculations).
Since both ${\mathbm S}_{\rm DC}$ and ${\mathbm S}_{\rm LC}$
eliminate two entries of the observation vector, 
the restricted DC and LC channel calculations 
map five entries of the observation vector onto five parameters
of the $\widehat{\mathbm p}$ vector.
For the restricted DC channel, 
the restricted observation vector $\mathbm y_{\rm DC}$,
the restricted uncertainty vector $\mathbm sigma_{\rm DC}$,
the Jacobian $\mathbm A_{\rm DC}$,
the covariance matrix $\mathbm V_{\rm DC}$ of observables, 
the correlation matrix $\mathbm R_{\rm DC}$ of observables, 
and the covariance matrix $\mathbm C_{\rm DC}$ of the parameters,
are obtained as follows,
\begin{subequations}
\label{DCrestricted}
\begin{align}
\mathbm y_{\rm DC} =& \; {\mathbm S}_{\rm DC} \, \overline{\mathbm y} 
= (0.12,\, 0.07, 0.01,\, 0.20,\, 0.01)^T \,,
\\
\mathbm \sigma_{\rm DC} =& \; {\mathbm S}_{\rm DC}  \,
\boldsymbol{\sigma}_{\overline{\boldsymbol{y}}} = 
(0.01,\, 0.01, \, 0.10,\, 0.02,\, 0.01)^T \,,
\\
\mathbm A_{\rm DC} =& \; {\mathbm S}_{\rm DC} \, \mathbm A \,,
\\
\mathbm V_{\rm DC} =& \; {\mathbm S}_{\rm DC} \, \mathbm V \, 
({\mathbm S}_{\rm DC})^T \,,
\\
\mathbm R_{\rm DC} =& \; {\mathbm S}_{\rm DC} \, \mathbm R \, 
({\mathbm S}_{\rm DC})^T \,,
\\
\mathbm C_{\rm DC} =& \; [ ({\mathbm A}_{\rm DC})^T \, 
\mathbm V_{\rm DC} \, ({\mathbm A}_{\rm DC}) ]^{-1} \,.
\end{align}
\end{subequations}
The fitted model, for the restricted DC channel, is obtained 
from the relation
\begin{equation}
\widehat{\mathbm p}_{\rm DC} = 
\mathbm C_{\rm DC} \, ( \mathbm A_{\rm DC} )^T \,
( \mathbm V_{\rm DC} )^{-1} \, \mathbm y_{\rm DC} 
= (0.0088, 0, 0.0214, 0.1989, 1.4286)^T \,.
\end{equation}
The uncertainties of the parameters 
of the restricted fitted model are
\begin{equation}
{\mathbm u}(\widehat{\mathbm p}_{\rm DC}) =
\sqrt{ {\rm diag}( \mathbm C_{\rm DC} ) }
= (0.1956, 0.0975, 0.0865, 0.02046, 1.4286)^T \,.
\end{equation}
The result of the restricted DC model for the 
gravitational constant is thus numerically
a bit lower than for the full model
[see Eq.~\eqref{Gresult}],
\begin{equation}
\label{GDCresult}
\mathcal G_{\rm DC} = 0.0088 \pm 0.1956 \,,
\end{equation}
For the restricted LC channel, 
formulas analogous to Eq.~\eqref{DCrestricted} apply,
for the restricted observation vector $\mathbm y_{\rm LC}$,
the restricted uncertainty vector $\mathbm sigma_{\rm LC}$,
the Jacobian $\mathbm A_{\rm LC}$,
the covariance matrix $\mathbm V_{\rm LC}$ of observables, 
the correlation matrix $\mathbm R_{\rm LC}$ of observables, 
and the covariance matrix $\mathbm C_{\rm LC}$ of the parameters,
\begin{subequations}
\label{LCrestricted}
\begin{align}
\mathbm y_{\rm LC} =& \; {\mathbm S}_{\rm LC} \, \overline{\mathbm y}
= (0.14,\, 0.09,\, 0.01,\, 0.20,\, 0.01)^T \,,
\\
\mathbm \sigma_{\rm DC} =& \; {\mathbm S}_{\rm LC}  \,
\boldsymbol{\sigma}_{\overline{\boldsymbol{y}}} =
(0.03, \, 0.03,\, 0.10,\, 0.02,\, 0.01)^T \,,
\\
\mathbm y_{\rm LC} =& \; {\mathbm S}_{\rm LC} \,
\overline{\mathbm y} \,,
\\
\mathbm A_{\rm LC} =& \; {\mathbm S}_{\rm LC} \, \mathbm A \,,
\\
\mathbm V_{\rm LC} =& \; {\mathbm S}_{\rm LC} \, \mathbm V \, 
({\mathbm S}_{\rm LC})^T \,,
\\
\mathbm R_{\rm LC} =& \; {\mathbm S}_{\rm LC} \, \mathbm R \, 
({\mathbm S}_{\rm LC})^T \,,
\\
\mathbm C_{\rm LC} =& \; [ ({\mathbm A}_{\rm LC})^T \, 
\mathbm V_{\rm LC} \, ({\mathbm A}_{\rm LC}) ]^{-1} \,.
\end{align}
\end{subequations}
The fitted model, for the restricted LC channel, is obtained 
from the relation
\begin{equation}
\widehat{\mathbm p}_{\rm LC} = 
\mathbm C_{\rm LC} \, ( \mathbm A_{\rm LC} )^T \,
( \mathbm V_{\rm LC} )^{-1} \, \mathbm y_{\rm LC} 
= (0.1460, 0, -0.0176, 0.2009, 1.4286)^T \,.
\end{equation}
The uncertainties of the parameters 
of the restriced fitted model are
\begin{equation}
{\mathbm u}(\widehat{\mathbm p}_{\rm LC}) =
\sqrt{ {\rm diag}( \mathbm C_{\rm LC} ) }
= (0.1129, 0.0975, 0.0270, 0.0200, 1.4286)^T \,.
\end{equation}
The result of the restricted LC model for the 
gravitational constant is thus numerically
a bit higher than for the full model
[see Eq.~\eqref{Gresult}],
\begin{equation}
\label{GLCresult}
\mathcal G_{\rm LC} = 0.1460 \pm 0.1129 \,.
\end{equation}

%
% Cross Covariance
%
\subsection{Cross Covariance}
\label{sec6F}

The difference between the 
values for the gravitational constant obtained
from the DC and LC channels raises the 
question of their mutual consistency.
In order to answer this question reliably,
we should choose a more astute {\em ansatz}
than a mere calculation of their
difference under addition of the 
squares of their uncertainties.
Namely, it is necessary to calculate the 
cross covariance between the DC and LC channels,
which takes into account those sources of 
uncertainty that are common to the DC and LC
channels. To this end, we define the operators
\begin{subequations}
\begin{align}
\mathbm L_{\rm DC} =& \; \mathbm C_{\rm DC} \, ( \mathbm A_{\rm DC} )^T \,
( \mathbm V_{\rm DC} )^{-1} \,,
\\
\mathbm L_{\rm LC} =& \; \mathbm C_{\rm LC} \, ( \mathbm A_{\rm LC} )^T \,
( \mathbm V_{\rm LC} )^{-1} \,,
\\
\mathbm V_{\rm DCLC} =& \; 
\mathbm S_{\rm DC} \, \mathbm V_{\overline{\mathbm y}} \,
( \mathbm S_{\rm LC} )^T \,,
\\
\mathbm C_{\rm DCLC} =& \; 
\mathbm L_{\rm DC} \, \mathbm V_{\rm DCLC} \,
( \mathbm L_{\rm LC} )^T \,,
\end{align}
\end{subequations}
The $\mathbm S$ operators are naturally identified as selection
operators, while the linear $\mathbm L$ operators connect
the parameter estimates with the observations [see Eq.~\eqref{Lconnect}].
The covariance matrix of the combined DC and LC channels
has the structure
\begin{equation}
\mathbm{\mathcal C}_{{\rm DC} \times {\rm LC}} =
\left( \begin{array}{cc}
\mathbm{C}_{\rm DC} & \mathbm C_{\rm DCLC} \\
(\mathbm C_{\rm DCLC})^T & \mathbm{C}_{\rm LC} 
\end{array}
\right) \,.
\end{equation}
This matrix is a square matrix. If 
the parameter vector $\widehat{\mathbm p}_{\rm DC}$
has $N_{\rm DC}$ entries, the 
the parameter vector $\widehat{\mathbm p}_{\rm LC}$
has $N_{\rm LC}$ entries, then
$\mathbm{\mathcal C}_{{\rm DC} \times {\rm LC}}$
is an
$(N_{\rm DC} + N_{\rm LC}) \times (N_{\rm DC} + N_{\rm LC})$
matrix. We remember that the 
parameter vectors $\widehat{\mathbm p}_{\rm DC}$
and $\widehat{\mathbm p}_{\rm DC}$ can have 
entries which correspond to the same physical 
quantity, such as the gravitational constant,
which is being determined using two different
methods. 

In order to put the concept of a cross covariance
into context, let us remember that 
cross covariances naturally appear in a number of 
engineering problems such as the 
aircraft-track problem studied in 
Refs.~\cite{BarShalom1981Correlation,BarShalomCampo1986,ChangSahaBarShalom1997},
where the optimized tracking of 
an aircraft trajectory using two different sensors
(radars) is being studied. 
We also refer to an application in the medical 
field described in Ref.~\cite{Kubacek2013SUR}, 
where the concept of a cross covariance leads
to additional insights.  However, let us continue 
our brief detour and discuss aircraft
tracking in the formalism of Ref.~\cite{ChangSahaBarShalom1997},
notably, the so-called fusion of aircraft trajectory tracking..
In the terminology of 
Ref.~\cite{ChangSahaBarShalom1997}, fusion does
not refer to nuclear fusion but the statistical combination of
two estimates of the same physical state.
We assume that sensors (radars) $i$ and $j$ produce an
estimated target state, for example a vector containing position and
velocity, $\widehat{\bm x}_i$ and $\widehat{\bm x}_j$.
The covariance of the track difference is obtained as 
$\Var(\widehat{\bm x}_j- \widehat{\bm x}_i)
= \Var(\widehat{\bm x}_j) + \Var(\widehat{\bm x}_i)
- \Cov(\widehat{\bm x}_i, \widehat{\bm x}_j)
- \Cov(\widehat{\bm x}_j, \widehat{\bm x}_i)$,
where ``$\Var$'' refers to the covariance
matrix of an individual track and ``$\Cov$'' refers to the 
cross covariance matrix of the tracks recorded by 
sensors $i$ and $j$.
With the identification
$\bm P_i = \Var(\widehat{\bm x}_i)$,
$\bm P_j = \Var(\widehat{\bm x}_i)$,
$\bm P_{ij} = \Cov(\widehat{\bm x}_i, \widehat{\bm x}_j)$ and
$\bm P_{ji} = \Cov(\widehat{\bm x}_j, \widehat{\bm x}_i)$,
the covariance matrix 
$\Var(\widehat{\bm x}_j- \widehat{\bm x}_i)$ 
of the difference of the tracks $i$ and $j$
enters the denominator in 
Eq.~(1) of Ref.~\cite{ChangSahaBarShalom1997},
which describes the update of the best estimate
for the tracking from sensor $i$ under the 
inclusion of the 
additional information from the added sensor $j$.

Let us now return to the comparison of the 
DC and LC channels in our proposed experiment.
Much in analogy to the aircraft track difference
in the discussed engineering application,
the covariance of the parameter-vector difference 
between the DC and LC channels is
\begin{equation}
  \Var(\widehat{\mathbm p}_{\rm DC} - \widehat{\mathbm p}_{\rm DC})
  = \mathbm C_{\rm DC} + \mathbm C_{\rm LC} \\
  - \mathbm C_{\rm DCLC} - (\mathbm C_{\rm DCLC})^T \,.
  \label{eq:dc-lc-difference-covariance}
\end{equation}
For the gravitational constant,
one remembers that the entry in the first row and first column of 
$\mathbm{C}_{\rm DC}$ pertains to the uncertainty square
of $\mathcal G_{\rm DC}$,
while the entry in the first row and first column of 
$\mathbm{C}_{\rm LC}$ pertains to the uncertainty square
of $\mathcal G_{\rm LC}$.  This implies that the
the entry in the first row and first column of 
$\mathbm{C}_{\rm DCLC}$
pertains to the cross covariance of 
of $\mathcal G_{\rm DC}$ of $\mathcal G_{\rm LC}$.
Hence, the uncertainty 
$u^2( \mathcal G_{{\rm DC}-{\rm LC}} ) = 
u^2( \mathcal G_{\rm DC} - \mathcal G_{\rm LC} )$ is 
\begin{equation}
u^2( \mathcal G_{{\rm DC}-{\rm LC}} ) = 
(\mathbm{C}_{\rm DC})_{11} + (\mathbm{C}_{\rm LC})_{11} -
2 (\mathbm{C}_{\rm DCLC})_{11} \\
= ( 0.03826 ) + ( 0.01274 ) - 2 \times ( 0.0097 ) 
= 0.0317 \,,
\end{equation}
so that the uncertainty assigned to the difference
is obtained as 
\begin{equation}
u( \mathcal G_{{\rm DC}-{\rm LC}} ) = 
\sqrt{ u^2( \mathcal G_{{\rm DC}-{\rm LC}} ) } = 0.1780 \,,
\end{equation}
and
\begin{equation}
\mathcal G_{\rm DC} - \mathcal G_{\rm LC} = 
-0.1372 \pm 0.1780 \,.
\end{equation}
For our hypothetical model, this result indicates
consistency between the DC and LC channels at the 
level of the experimental uncertainties. 

The comparison between the DC and $\omega$ science channels therefore
provides considerably more than a simple cross-check of the measured
value of \(G\). It constitutes an independent statistical hypothesis
test of the completeness of the entire observation model. 
In this sense, the covariance analysis not only predicts the uncertainty of
the combined estimate for $G$, but also predicts quantitatively
the level of agreement that should be observed between two distinct
physical realizations of the experiment. 
This internal consistency
test is one of the principal strengths of the proposed dual-channel
measurement strategy.

\section{Mission Parameters}
\label{sec7}

\subsection{LISA Power Spectral Density}
\label{sec7A}

One of the most important considerations for the 
final expected accuracy for $G$ (and for other mission
parameters) concerns the 
achievable acceleration-noise level,
which, for the LISA Pathfinder, can be found in Sec.~III A of
Ref.~\cite{Armano2016PRL}.
It is usually expressed as the square root of a power
spectral density (PSD), {\em i.e.}, as an amplitude spectral density (ASD). In
our estimates, we shall employ the first LISA Pathfinder results from
Ref.~\cite{Armano2016PRL} for the white differential acceleration ASD,
acknowledging that later mission results have meanwhile reached even lower
values~\cite{Armano2018PRL,Armano2024InDepth}.
We follow the analysis presented in Ref.~\cite{Armano2016PRL}
and denote the (one-sided) power spectral density of the residual
differential acceleration noise by $S_a$.

For clarification, we should remark that 
the word ``power'' in \emph{power spectral density} refers to
refers to a mean-square strength of a stationary random process,
in this case, to the acceleration noise and follows
standard conventions used in random-signal analysis.
The ``power'' is not measured in units of Watts,
in this case, but rather, if $a(t)$ is an acceleration noise,
then $[S_a] = \frac{({\rm m\,s^{-2}})^2}{\rm Hz}$.
Furthermore, remarks from Ref.~\cite{Armano2018PRL} clarify that 
the unit of ``Hz'' should rather be understood 
as ``rad/s'' (radians per second), and that the 
actual noise at frequency $f$ is obtained
by multiplying the results for $S_a$ by $2 \pi \, f$.
It may be permitted to 
rematk that it has recently been advocated 
in Ref.~\cite{MohrEtAl2022} to address a corresponding
possible source of misunderstandings by 
redefining the SI unit system, acknowledging that 
an angle is a physical quantity with a dimension.

In order to gauge the situation, it is 
useful to remember the so-called one-sided,
and two-sided, conventions for the calculation
of the power spectral density. In general,
the relation between autocorrelation and spectral density is the
Wiener--Khintchine theorem~\cite{Wiener1930,Khintchine1934}
Let $n(t)$ be a real, zero-mean, stationary random acceleration with
autocorrelation
\begin{equation}
R_n(\tau)=\mathbb E[n(t) \, n(t+\tau)] \,. 
\end{equation}
One defines the two-sided PSD $S_n^{(2)}(f)$
at frequency $f$ as an integral over $\tau$,
with $\tau$ ranging from $-\infty$ to $+\infty$,

\begin{equation}
S_n^{(2)}(f) =
\int_{-\infty}^{+\infty}
R_n(\tau)e^{-i2\pi f\tau}\, \dd \tau \,.
\end{equation}
Then, for $\tau = 0$, one obtains
\begin{equation}
\mathbb E[n^2(t)] = R_n(0) =
\int_{-\infty}^{+\infty} S_n^{(2)}(f)\,\dd f\,.
\end{equation}
For a real rather than complex function $R_n(\tau)$,
one obtains the result $S_n^{(2)}(-f)=S_n^{(2)}(f)$.
The positive- and negative-frequency halves therefore contain equal
noise power. This leads to the notion that it will
be possible to define a one-sided PSD 
that folds the negative-frequency contribution onto the
positive-frequency axis. 
So, if one defines the one-sided PSD 
$S_n^{(1)}(f)$ with a manifestly positive
argument $f > 0$ according to 
\begin{equation}
S_n^{(1)}(f) = 2 \, S_n^{(2)}(f) \,,
\end{equation}
then $\mathbb E[n^2(t)]$ can be expressed as
a ``one-sided'' integral,
\begin{equation}
\mathbb E[n^2(t)] = \int_0^{+\infty}S_n^{(1)}(f)\,\dd f\,.
\end{equation}

The total variance of $n(t)$ 
is unchanged; only the bookkeeping convention has
changed. Hence, one can express the
one-sided amplitude spectral density
${\rm ASD}(f)$ as
\begin{equation}
{\rm ASD}(f) = \left[S_n^{(1)}(f)\right]^{1/2}
= \sqrt{2}\, \left[S_n^{(2)}(f)\right]^{1/2} \,. 
\end{equation}
The one-sided PSD convention is standard in gravitational-wave data
analysis \cite{Allen2012Findchirp,Armano2016PRL}.

We now study the white-noise correlation in the one-sided convention,
and assume that, over the narrow frequency interval relevant 
to the lock-in measurement (ideally, one would induce
source-mass oscillations at a single, well-defined
frequency), the one-sided acceleration PSD is approximately
constant, $S_a^{(1)}(f)\simeq S_a$. Then,
inverting the Fourier transform in the one-sided
conventions, one has 
\begin{equation}
R_n(\tau)=\mathbb E[n(t) \, n(t+\tau)] 
= \int_0^\infty \frac{ \dd \omega }{2 \pi} \,
\ee^{-\ii \omega \tau} \, S_a = \frac{S_a}{2}\,\delta(\tau) \,,
\end{equation}
where the integration region for the Dirac-$\delta$ 
has been centered at $\omega = 0$ and a symmetric
representation of the Dirac-$\delta$ distribution about its
center has been assumed. For ideal white noise,
\begin{equation}
\mathbb E[n(t)n(t')] =
\frac{S_a}{2}\,\delta(t - t') \,.
\end{equation}
The factor \(1/2\) is therefore a convention factor associated with the
one-sided PSD; it is not an additional physical noise contribution.

One now considers, for reference, 
resonant driving of the test masses at angular frequency $\omega$,
\begin{equation}
a(t)=A\cos(\omega t)+n(t),
\end{equation}
measured for a duration \(T\), with \(T\) containing an integer number
of modulation periods.  The phase-sensitive estimator is
easily obtained as
\begin{equation}
\widehat A = A + 
\int_0^T n(t)\cos(\omega t)\,\dd t .
\end{equation}
Its variance is
\begin{equation}
\sigma_A^2 = \mathbb E[(\delta A)^2]
= \frac{4}{T^2} \int_0^T \dd t \, \int_0^T \dd t'
\cos(\omega t) \, \cos(\omega t')
= \frac{4}{T^2} \frac{S_a}{2} \frac{T}{2} = \frac{S_a}{T} \,.
\end{equation}
Therefore, the uncertainty of the acceleration
of the test masses due to white noise 
can be estimated as
\begin{equation}
\label{sigmaA}
\sigma_A=\frac{S_a^{1/2}}{\sqrt{T}}.
\end{equation}

\subsection{Optimum Source-Mass Scale}
\label{sec7B}
 
In principle, one might think that the source mass should 
be increased as far as possible in order to enhance the 
gravitational signal observed in the experiment.
However, increasing the source mass beyond the point where the
statistical uncertainty becomes significantly smaller than 
other systematic uncertainties, 
will yield progressively smaller improvements in the
overall determination of $G$.
Furthermore, very large source masses substantially increase the
engineering complexity of the mission.
Examples include structural support,
source actuation,
preflight dimensional metrology,
multipole characterization,
and thermal control (including uniform thermal stabilization).

Our preceding analysis therefore suggests that an optimum source-mass
range should exist rather than a monotonic preference for the largest
possible masses.
A preliminary estimate of the useful source-mass scale may be obtained
by comparing the statistical uncertainty of the gravitational signal
with an assumed residual systematic-uncertainty floor.

We proceed as follows. First, as demonstrated in Sec.~\ref{sec7A},
using the Wiener--Khintchine theorem,
one can derive, for a cosinusoidal coherently fitted signal observed over an
effective integration time $T$,
the standard uncertainty of the fitted
amplitude as given in Eq.~\eqref{sigmaA}.
The signal acceleration $A_G$ can be obtained 
from Eq.~\eqref{eq:A0} and Fig.~\ref{fig1} as
follows [see also Eq.~\eqref{eq:A0}],
\begin{equation} 
A_0 = \frac{8GMd_0s}{(d_0^2-s^2)^2} \simeq A_G \equiv \frac{8GMd_0s}{s^3}.
\end{equation} 
For $s = 1\,$m and $d_0 = 10$\,m, one obtains
\begin{equation}
\label{AG}
A_G = 5.448\times10^{-10}
\left(\frac{M}{1~{\rm t}}\right)
{\rm m\,s^{-2}}.
\end{equation}
The corresponding fractional statistical uncertainty in \(G\) is
\begin{equation}
u_{\rm stat}(G) = \frac{\sigma_A}{A_G}
= \frac{S_a^{1/2}} {A_G\sqrt{T}}.
\end{equation}
For the numerical estimate, take the representative acceleration-noise
level from Ref.~\cite{Armano2016PRL},
\begin{equation}
\label{LISAnoise}
S_a^{1/2} = 5.2~{\rm fm}\,{\rm s}^{-2}/\sqrt{\rm Hz},
\end{equation}
which is comparable to the differential free-fall performance demonstrated by
LISA Pathfinder in the millihertz band
(see also Appendix~\ref{appa}). 
This value is used only as a
reference noise scale and should not be interpreted as a guaranteed
performance level for the proposed experiment. 
For reference, we point out that in Ref.~\cite{Armano2016PRL},
a value of $(5.2\pm0.1)~{\rm fm}\,{\rm s}^{-2}/\sqrt{\rm Hz}$ was
reported in a frequency band 
between approximately $0.7$ and $20~{\rm mHz}$.
One notes that this result has not been assigned 
an equation number in Ref.~\cite{Armano2016PRL},
but can be found in the abstract and in Sec.~III A
of Ref.~\cite{Armano2016PRL}.
Substitution gives
\begin{equation}
u_{\rm stat}(G) = 0.0325~{\rm ppm}
\left(\frac{1~{\rm t}}{M}\right)
\left(\frac{1~{\rm day}}{T}\right)^{1/2}.
\end{equation}
For an effective integration time of $30$~days,
\begin{equation}
u_{\rm stat}(G) = 0.00593~{\rm ppm}
\left(\frac{1~{\rm t}}{M}\right)
\left(\frac{30~{\rm days}}{T}\right)^{1/2}.
\end{equation}
The ideal statistical uncertainties are therefore
\begin{subequations}
\label{ideal}
\begin{align}
u_{\rm stat}(G) =& \; 0.593~{\rm ppm} 
& \qquad & 
(M = 10~{\rm kg}, T = 30~{\rm days}) \,, \\
u_{\rm stat}(G) =& \; 0.0593~{\rm ppm} 
& \qquad &
(M = 100~{\rm kg}, T = 30~{\rm days}) \,, \\
u_{\rm stat}(G) =& \; 5.93 \times 10^{-9}
& \qquad &
(M = 1000~{\rm kg}, T = 30~{\rm days}) \,.
\end{align}
\end{subequations}
The ideal source mass range therefore depends
on the target accuracy for the experiment 
and on the launch capabilities of modern 
spacecraft. 

For the {\sc Ariane~5} of the European Space Agency (ESA),
the payload capacity to GTO is $\approx 10~{\rm t}$
(see Ref.~\cite{ESA_Ariane5ECA}),
while, for the {\sc Ariane}, a slight improvement 
to approximately $11.5~{\rm t}$ to GTO 
is planned according to the ESA design 
documents for the four-booster Ariane~6
configuration~\cite{ESA_Ariane6Overview}.
The {\sc Saturn V} could carry $44.9~{\rm t}$
to a trans-lunar injection (TLI) orbit
while for a low--Earth orbit, 
the payload was given as $118.8~{\rm t}$
(see Ref.~\cite{SumrallCreech2009AresV}).
A standard GTO payload was not
a normal published mission rating;
it would have been larger than the
TLI payload but smaller than the low--Earth orbit 
payload and would require a
mission-specific trajectory calculation.
For the {\sc NASA Artemis SLS Block~1}
of the National Aeronautics and Space Administration (NASA),
the payload capacity to GTO is $\approx 25$--$35~{\rm t}$,
while for the {\sc NASA Artemis SLS Block~2},
the design payload is $55.0~{\rm t}$
(see Ref.~\cite{NASA_SLS_MPG2018}).
For reference, a GTO is an elliptical orbit whose apogee lies near the
radius of the geostationary Earth orbit (GEO), while the spacecraft
itself performs the orbit raising burn, 
and the final circularization burn at the apogee
of the elliptical GTO, and the 
inclination correction into GEO using its own propulsion system.
Finally, the planning for the {\sc SpaceX Starship}
calls for payloads in excess of $100\,$t (metric tonnes)
to be lifted, first, into a low Earth orbit, and,
upon refueling, even onto the lunar surface
(see Refs.~\cite{Musk2018Multiplanetary,%
SpaceXStarshipGuide2020,SpaceXMoonRefilling}).
Since essentially all commercial geostationary spacecraft are launched
via GTO, payload capabilities quoted for GTO constitute the standard
figure of merit adopted by launch providers,
and are used as the basis for our estimates
in the current investigation. For the present mission,
the spacecraft would likewise be assumed to 
employ an onboard propulsion system for
the final transfer from GTO to the operational orbit.

Assuming an ideal realization of the achievable 
statistical uncertainty (and a negligible error
from the calibration of the source masses), 
for a target accuracy of 
\begin{equation}
10^{-9} < u_{\rm stat}(G) < 10^{-8} \,,
\end{equation}
an appropriate source mass range can be estimated as
\begin{equation}
0.6~{\rm t} \lesssim M \lesssim 6~{\rm t} \,.
\end{equation}

%
% Candidate Geostationary Location
%
\subsection{Candidate Geostationary Location}
\label{sec7C}

While a high Earth orbit with an orbital period of approximately
two days could be considered because of the relatively quiet gravitational
environment, a geostationary orbit (GEO) offers 
overwhelming operational advantages.
Specifically, a spacecraft in geostationary orbit provides
essentially continuous communication with a single ground station,
continuous telemetry and health monitoring,
uninterrupted science operations,
a nearly constant thermal environment,
mature orbit-determination techniques developed over decades of
satellite operations,
and continuous solar power except during the short eclipse seasons.

These features are particularly attractive for a precision experiment
requiring long uninterrupted integration times.
The geostationary belt is approximately
\begin{equation}
R_{\rm GEO}=42\,164~{\rm km}
\end{equation}
from the Earth's center and extends over all longitudes.
Although several hundred satellites occupy geostationary and
geosynchronous orbits, the distribution is far from uniform.
Commercial communications satellites cluster preferentially above
densely populated continents.
An obvious, particularly attractive candidate location 
would be at the longitude of 
\begin{equation}
\lambda \approx 155^\circ\ {\rm W},
\end{equation}
over the equator,
$2400~{\rm km}$ south of the Hawaiian Islands, over the central Pacific Ocean.
The corresponding subsatellite point, of course,
is at $0^\circ$ latitude and a longitude of $155^\circ \, {\rm W}$.
This longitude lies in one of the comparatively quiet portions of the
geostationary belt while still retaining all operational advantages of
a GEO.

For the indicated location south of the Hawaiian Islands,
a close neighbor on the GEO would be the current GOES-West weather satellite,
whose geostationary location is near $137^\circ{\rm W}$.  The angular separation
is therefore approximately $\Delta\lambda \approx 18^\circ$, and the
corresponding arc length along the geostationary ring is $(R_{\rm GEO}) \times
(\Delta\lambda) \times \frac{\pi}{180} = 13\,200~{\rm km}$, which places the proposed
spacecraft many thousands of kilometers away from one of the nearest major
operational satellites.  At such distances, the direct gravitational influence
of neighboring spacecraft is negligible, and the local spacecraft traffic is
considerably lighter than over the principal commercial longitudes.

\section{Conclusions}
\label{sec8}

We have presented a general concept of a space-based measurement of the
gravitational constant $G$; some elements of the proposal 
are very specific (for example, mission parameters),
while other elements (for example, the use of cross-covariances
in order to analyze the statistics of different 
experimental channels) could be useful in a much
more general context. Unlike many traditional measurements of the
gravitational constant, whose principal objective is to maximize the
gravitational signal, the present proposal makes heavy use of the 
method of covariances.
The attainable uncertainty of the adjusted value of Newton's
gravitational constant is determined not only by the statistical precision of
the individual observations but also by the extent to which the desired
parameter, namely, $G$,
may be distinguished from all remaining physical perturbations.  We
are thus willing to increase the number of nuisance parameters, provided we
simultaneously increase the number of observables even more,
leading to an overdetermined system.  That is a very
different experimental philosophy, which is also applied to the recent CODATA
adjustments~\cite{Tiesinga2021CODATA,Tiesinga2021CODATA}.

The proposed mission observes the controlled trajectories of macroscopic test
masses in an engineered gravitational field, while the dominant
non-gravitational couplings are measured through auxiliary channels.  The
concept is technically ambitious but rests on several mature foundations.  The
SEE established the scientific value of space-based relative trajectories for
measuring gravitational
parameters~\cite{SandersDeeds1992,AlexeevEtAl1994,Sanders1996CommentsSEE,%
SandersEtAl1999,SandersEtAl2000,AlexeevEtAl2000SEE,Alexeev2001SEE,%
Sanders2010SEE,Melnikov2016SEE}, while LISA Pathfinder demonstrated that
free-falling metallic test masses, capacitive sensing, electrostatic actuation,
charge management, and interferometric displacement readout can operate
together at extraordinary
sensitivity~\cite{Armano2016PRL,Armano2018PRL,Armano2024Actuation}.  
We propose to use the laser-interferometric displacement readout techniques in
conjunction with two different experimental channels, a direct channel with an
approximately uniform acceleration of the test masses, and a lock-in channel
with a periodic modulation of the source-mass positions.

In our proposed geometry, the additional introduction of a third test mass
permits the construction of two symmetry-adapted observables.  The signal
coordinate responds predominantly to the desired gravitational signal produced
by the movable source masses.  The null coordinate responds primarily to
common-mode perturbations, including residual spacecraft deformation, thermal
expansion, optical reference motion, and other environmental disturbances.  The
null coordinate therefore serves not merely as a diagnostic but as an
additional observable constraining nuisance parameters within the least-squares
adjustment.  A rotation of the source masses addresses multipole effects.  The
comparison of signals from the DC and LC channels leads to an important
consistency check for the final value of $G$ (see Sec.~\ref{sec6F}).
One notes that the separation of the experimental signal from background
effect is particularly challenging for the gravitational interaction 
which cannot be shielded, not even by the presence of antiparticles, 
which, as can be shown (see Refs.~\cite{JeNo2013pra,Je2019ijmpa,Je2020physics}),
undergo the same gravitational acceleration as their corresponding 
particles, within both the quantum as well as classical theories.

The detached-laboratory architecture outlined in Fig.~\ref{fig2} permits a
cleaner geometry, in the sense that heavier source masses may be placed farther
away, reducing the influence of multipoles and effective-center errors while
preserving the interferometric signal.  The philosophy is to use additional
resources on reducing model dependence rather than maximizing amplitude.
Separating the experiment from the service spacecraft before measurement
strongly suppresses spacecraft self-gravity, vibration, and thermal radiation
without imposing a meaningful communication delay. The service spacecraft can
remain hundreds of meters away and communicate with microsecond latency, while
the laboratory spacecraft undergoes long quiet free-flight intervals. 

A central element of the proposal is the covariance analysis, which can be used
in order to extract optimum values of $G$ from the combined DC and LC channels,
and which provides important cross-checks. In particular, The consistency of
the the optimum parameters $\widehat{\bm p}_{\rm DC}$ and $\widehat{\bm p}_{\rm
LC}$ obtained from the DC and LC channels serves as an important cross-check
against any inconsistencies of the evaluation, and of the underlying analytic
model.  In order to perform this cross-check meaningfully, one needs to use the
method of cross covariances (see Sec.~\ref{sec3F}).  The covariance analysis is
made easier by the approximate reflection symmetry of the proposed apparatus,
which provides a series of first-order selection rules governing the coupling
between physical parameters and experimental observables (as outlined in
Sec.~\ref{sec3}). Several Jacobian elements therefore vanish identically to
first order, while others are strongly suppressed (see also Sec.~\ref{sec6D}
for an illustrative example).  These selection rules are independent of the
numerical values of the physical parameters and arise solely from the symmetry
of the experimental geometry.  Rather than relying exclusively upon numerical
optimization, the experiment exploits symmetry itself to suppress parameter
correlations.

Finally, one might ask to which extent the lock-in channel could be used for
terrestrial measurements; this question is being addressed in
Appendix~\ref{appa} with a nuanced answer, which highlights the advantages of
the combination of the lock-in method with the space-based environment.  The
extent to which alternative theories of gravity, which lead to small deviations
from the Newtonian laws, could be tested in the space-based apparatus, is being
discussed in Appendix~\ref{appb}. 

In summary, the present work has developed a covariance-based formulation of a
space-based determination of Newton's gravitational constant.  While our
proposed apparatus certainly does not constitute {\rm the only} possible design
philosophy, we hope that we were able to demonstrate that it has a number of
tangible advantages. Unlike conventional uncertainty analyses, which primarily
propagate independently estimated systematic corrections into the final
uncertainty budget, the present formulation regards many systematic effects as
explicit parameters of a single generalized least-squares adjustment.  The
resulting estimation problem has been formulated in terms of a set of
observables together with an associated parameter vector
describing both Newton's gravitational constant and the principal ``nuisance''
parameters of the experiment, where ``nuisance'' should be understood
{\em cum grano salis}. The method of covariances
turns the ``nuisance parameters'' into ``auxiliary parameters''
which help in improving the final result.
If we take the optimistic view that apparatus-related
effects such as thermal effects can be brought under control, then the ultimate
limits of accuracy of our proposal lie in the mass calibration of the source
masses, and in the white-noise contribution discussed in Secs.~\ref{sec7A}
and~\ref{sec7B} [see also Eq.~\eqref{ideal}].  Finally, let us also express the
hope that three essential ingredients of our proposal, namely, the dual-channel
lock-in configuration, the method of (cross-)covariances, and the introduction
of null coordinates could be useful in a wider context, perhaps, even, in
future Earth-based experiments, especially when modified and adapted to
different experimental geometries.
Current discrepancies among different determinations of $G$
provide a compelling reason to pursue independent methods with different
systematic-error structures~\cite{Quinn2000MeasuringBigG,Quinn2014DontStopBigG}.
The present concept is offered in that spirit.

%
% Acknowledgments
%
\section*{Acknowledgments}

The author acknowledges insightful and elucidating
conversations with P.~J.~Mohr and S.~Schlamminger,
during an extended visit 
at the National Institute of Standards and Technology (NIST).
These discussions were a great inspiration for this paper,
and the main ideas for this work originated from 
discussions at NIST about possible advances in the 
determination of fundamental physical constants,
in connection with the CODATA Task Group on 
Fundamental Constants.
Support from the National Science Foundation
(Grant PHY--2513220) is gratefully acknowledged.
Portions of the text were partially created 
and revised with assistance from {\tt gpt-5.1-high}.
Also, assistance from {\tt gpt-5.1-high} was
helpful in the derivation of the precise form of
the symmetry relations reported in Sec.~\ref{sec3}.
The author independently performed
and verified all calculations, interpretations, and scientific
conclusions presented in this paper,
and assumes full and complete responsibility for the work.

%
% Data Availability
%
\section*{Data Availability}

The conclusions of the article are supported by the analytic formulas given in
the manuscript. No experimental data were created or analyzed in the purely
theoretical studies. In particular, the numerical sample data which
constitute the basis for the illustrative example described in Sec.~VI are
contained in the article itself [see Eqs.~(6.19)---(6.23)]. 

%
% Data Availability
%
\section*{Conflicts of Interest}

The author declares no conflict of interest.

%
% Ethics Statement
%
\section*{Ethics Statement}

This study does not involve
human subjects, human data or tissue, or animals. 
No ethics questions have arisen.

\appendix

\section{Lock--In and Terrestrial Measurements}
\label{appa}

The use of a periodically modulated source configuration 
constitutes a general precision-measurement strategy,
and there are important terrestrial precedents in which rotating or
periodically reconfigured attractor masses move the gravitational signal
away from zero frequency~\cite{GundlachEtAl1996,GundlachMerkowitz2000,Hoyle2004ISL,%
Tu2007DualMod,Tan2016DualMod}.
A continuous attractor rotation was explicitly used to reduce background
noise in Refs.~\cite{GundlachEtAl1996,GundlachMerkowitz2000}. Closely related modulation
strategies have also been employed in precision tests of the
inverse-square law, including experiments with rotating multipole
attractors~\cite{Hoyle2004ISL,Tu2007DualMod,Tan2016DualMod}.

The present discussion therefore does not claim novelty for source-mass
modulation itself.  Rather, it is useful to ask what advantages and
limitations arise if the static and modulated channels are deliberately
combined in a terrestrial determination of $G$.
Consider, schematically, a source--test-mass separation
\begin{equation}
d(t)=d_0+\delta d\cos(\omega t),
\qquad
\delta d\ll d_0 .
\end{equation}
For an acceleration \(a(d)\), expansion about \(d_0\) gives
\begin{equation}
a[d(t)] \simeq a(d_0) + \delta d 
\left.  \frac{\partial a}{\partial d} \right|_{d_0}
\cos(\omega t)
+\mathcal O(\delta d^2).
\end{equation}
Thus a terrestrial experiment may naturally define two complementary
observables,
\begin{equation}
y_{\rm DC}\sim a(d_0), \qquad
y_{\omega}\sim \delta d\,
\left.  \frac{\partial a}{\partial d} \right|_{d_0}.
\end{equation}
The first probes the static gravitational response, whereas the second
probes the response synchronous with the controlled source motion.
For Newtonian point masses,
$a(d)=\frac{GM}{d^2}$
and hence
$y_\omega \propto -\,2\,\frac{GM}{d_0^3}\,\delta d$.
For an extended source distribution the corresponding quantities are
obtained by integrating over the measured density distribution.

A continuous rotation of an attractor is mathematically analogous.
If the source geometry possesses an \(m\)-fold azimuthal symmetry and
the attractor rotates at angular frequency \(\Omega\), the gravitational
signal can appear predominantly at the harmonic \(m\Omega\).  This
principle has been exploited directly in rotating-attractor torsion
balances \cite{Hoyle2004ISL,Tan2016DualMod}.

A principal advantage of the lock-in channel is spectral translation.
A purely static gravitational signal lies near zero frequency, where
many terrestrial instruments suffer from slow thermal drift, tilt,
creep, changes of laboratory conditions, and low-frequency noise.
Periodic source motion moves the desired signal to a chosen frequency
\(\omega\), so that the relevant quadratures may be extracted by
phase-sensitive regression, much in analogy
to the approach discussed here
for the space-based apparatus.
Noise that is not coherent with the source modulation is strongly
suppressed by long coherent integration.  Continuous attractor rotation
was used for precisely this general purpose in the measurement of
Ref.~\cite{GundlachMerkowitz2000}.

Another advantage is geometrical discrimination.  The DC and
lock-in responses have different dependences on source distance,
multipole moments, and possible deviations from Newtonian gravity.
Terrestrial atom-interferometric determinations of \(G\), although not
generally based on continuous sinusoidal modulation, illustrate the
value of deliberately changing nearby source-mass configurations and
measuring the resulting differential gravitational field
\cite{Lamporesi2008G,Rosi2014G}. 

The most important qualification is that lock-in detection rejects
\emph{incoherent} disturbances, but it does not reject disturbances
that are themselves synchronous with the source motion.  This is the
central challenge of a terrestrial LC channel.
Namely, if a source mass \(M_s\) is translated according to
$x_s(t)=\delta d\cos(\omega t)$,
the actuator must provide an inertial force of order
$F_{\rm act} \sim M_s\omega^2\delta d$.
The equal and opposite reaction force acts on the source support,
foundation, or laboratory structure.  Any resulting motion of the
detector reference frame at the same frequency can be indistinguishable
from the desired gravitational response unless it is independently
measured or rejected by symmetry.  This problem becomes increasingly
serious for large source masses or high modulation frequencies.
Source-synchronous disturbances include
mechanical feedthrough and seismic coupling,
tilt and elastic deformation,
and thermal effects, because the 
periodic actuator power can produce temperature variations synchronous
with the source motion.
Thermal expansion changes baselines and source
positions; temperature gradients can alter suspension properties,
optical paths, residual-gas forces, and radiative forces.  

On Earth, the source drive and the detector ultimately reside in a mechanically
connected environment, so actuator reaction forces can reach the detector
through the laboratory structure.  In the proposed space experiment, the test
masses are freely falling, and the enclosing laboratory can be designed so that
ordinary mechanical transmission from the source actuator to the test masses is
strongly suppressed.  In principle, source-mass modulation and phase-sensitive
detection are general metrological techniques and can improve terrestrial
measurements of $G$, particularly by moving the signal away from low-frequency
drift.  The space-based concept is distinguished not by the mere existence of a
modulated channel, but by combining DC and modulated gravitational measurements
with freely falling test masses, symmetric source geometry, null channels, and
a common covariance adjustment in an environment where mechanical coupling
between the moving sources and the test masses can be greatly suppressed.

\section{Tests of Non-Newtonian Gravity}
\label{appb}

A precision measurement of Newton's gravitational constant over
source--test-mass separations of order meters is simultaneously a test
of the distance dependence of gravity. Many extensions of the Standard
Model and of general relativity predict additional weak interactions
which, in the nonrelativistic limit, modify the Newtonian potential.
A standard model-independent parameterization is the Yukawa form
\cite{FischbachTalmadge1999,Adelberger2003ISL,Adelberger2009Torsion},
\begin{equation}
V(r) = -\frac{Gm_1m_2}{r}
  \left[ 1+\alpha \exp\!\left(-\frac{r}{\lambda}\right) \right],
\label{eq:yukawa-potential}
\end{equation}
where $m_1$ and $m_2$ are the two 
masses, $\alpha$ is the strength of the additional interaction relative
to Newtonian gravity and $\lambda$ parameterizes the 
Yukawa range. The corresponding mediator mass scale 
$m_\phi$ is
\begin{equation}
m_\phi c^2=\frac{\hbar c}{\lambda}
\simeq 1.97\times10^{-7}\ {\rm eV}
\left(\frac{1\,{\rm m}}{\lambda}\right) \,.
\end{equation}
The use of precision measurements of \(G\) themselves as
inverse-square-law tests has direct experimental precedent.
For the millimeter to meter range, examples include 
Refs.~\cite{Ke2021CombinedISL,Hoskins1985ISL,Kapner2007ISL,Yang2012ISL}. 
The present space experiment would therefore probe a complementary regime with
large source masses and long freely evolving test-mass trajectories at
meter-scale separations.

At a single fixed distance, a sufficiently long-range Yukawa
interaction may be partly degenerate with a small change of $G$.
For a spherically symmetric source mass $M$,
\begin{equation}
a(r) = \frac{GM}{r^2} \left[
    1+ \alpha \left(1+\frac{r}{\lambda}\right)
    \ee^{-r/\lambda} \right].
\end{equation}
At one value of \(r\), the term in square brackets can mimic a slightly
different effective gravitational constant. Measurements at several
source-mass separations break this degeneracy because the factor
$\left(1+\frac{r}{\lambda}\right) \exp(-r/\lambda)$
has a characteristic non-Newtonian distance dependence. This provides
an additional motivation for the two lock-in configurations
$y_{\rm LC}^{[1]}$ and $y_{\rm LC}^{[2]}$,
and, more generally, for a campaign employing several nominal
source-test-mass distances.

For a small radial source displacement
$r(t)=r_0+\delta r\cos(\omega t)$, with $\delta r\ll r_0$,
the leading synchronous acceleration is proportional to the local
spatial derivative of the force rather than the force itself.
For the Yukawa potential, the spatial derivative leads
to characteristic terms originating from the 
exponential factor, which could be separately 
detected in the lock-in channel.
For the actual two-source, three-test-mass apparatus, 
the distinction between the force and its local
spatial variation remains present and detectable in the integrated signals.

The exchange of a new ultralight scalar or vector boson can generate a
finite-range interaction of Yukawa form. A particularly interesting aspect is
that such forces may not only depend on the Yukawa parameter $\lambda$, but
also, on the chemical composition of the materials involved, or other
approximately conserved charges
\cite{Fischbach1986Eotvos,Adelberger2007ParticlePhysics,
Adelberger2009Torsion}. One could, in principle, probe for those forces using
test masses of different chemical composition.

Large or otherwise macroscopic extra dimensions can modify the
Newtonian force law. At distances comparable to a compactification
scale the potential can depart from the \(1/r\) form, while at larger
distances individual Kaluza--Klein contributions can appear as Yukawa
corrections. Representative theoretical motivations are given in
Refs.~\cite{ArkaniHamed1998Hierarchy,Antoniadis1998Millimeter}.

Screened scalar-tensor theories are especially interesting for a space
experiment. Chameleon models allow the effective mass and coupling of a
scalar degree of freedom to depend on the local matter 
environment. Various scenarios 
in this direction have been explored in 
Refs.~\cite{KhouryWeltman2004Space,HinterbichlerKhoury2010Symmetron,BurrageSakstein2018}. 
For a model-independent search, one may augment the observation model
according to
\begin{equation}
\overline{\mathbm y} = \mathbm f(\mathbm p)
+ \alpha\, \mathbm f_{\rm Y}(\mathbm p;\lambda) + \boldsymbol{\varepsilon},
\end{equation}
where $\mathbm f$ is the Newtonian model from Eq.~\eqref{full_observables}, and
$\mathbm f_{\rm Y}$ is the calculated response to a unit-strength
Yukawa interaction of range $\lambda$.


\begin{thebibliography}{82}
\bibitem [{REM()}]{REMARK}%
  \BibitemOpen
  \href@noop {} {}\bibinfo {howpublished} {A larger source mass can lead to
  stronger coupling to supports, and (unless extreme care is taken in regard to
  the geometry of the mass distribution) more difficult dimensional
  characterization and a more complicated multipole structure. Terrestrial
  experiments additionally contend with seismic motion, gravity gradients from
  moving groundwater and atmosphere, tilt, and temperature-dependent suspension
  properties.}\BibitemShut {Stop}%
\end{thebibliography}
\end{document}